# A New Photomechanical Molecular Switch

# Based on a Linear π-Conjugated System

*S. Lenfant\*, Y. Viero, C. Krzeminski, D. Vuillaume\**

Institute for Electronics Microelectronics and Nanotechnology (IEMN),

CNRS, Avenue Poincaré, Villeneuve d'Ascq, 59650, France.

*D. Demeter, I. Dobra, M. Oçafrain, P. Blanchard, J. Roncali*

MOLTECH-Anjou, CNRS, Univ. Angers, 2 Bd Lavoisier, Angers, 49045, France

*C. Van Dyck*

Chemistry Department, Northwestern University

2145 Sheridan Road, Evanston, Il 60208, USA

*J. Cornil*

Laboratory for Chemistry of Novel Materials, University of Mons,

20 Place du Parc, Mons, 7000, Belgium

ABSTRACT.

We report the electronic transport properties of a new photo-addressable molecular switch. The switching process relies on a new concept based on linear π-conjugated dynamic systems, in which the geometry and hence the electronic properties of an oligothiophene chain can be reversibly modified by the photochemical *trans-cis* isomerization of an azobenzene unit fixed in a lateral loop. Electron transport measurements through self-assembled monolayers on gold, contacted with eGaIn top contact, show switching with a conductance ratio up to $10^3$. *Ab initio* calculations have been used to identify the most energetically stable conformations of the molecular switch; the corresponding calculated conductances qualitatively explain the trend observed in the photo-switching experiments.

Electrical conductance switching associated to a change in the geometry of a molecule is a versatile phenomenon that is very useful to design and study nanoscale molecular switches and memories (from single molecule to monolayer). For example, conductance switching has been triggered by configurational *trans* to *cis* isomerization of azobenzene derivatives,[1-7] or to the closing/opening of carbon-carbon bond in diarylethene derivatives,[8-13] see a review on molecular switches in Refs. [14-16]. In the latter case, the change in conductance is primarily due to the breaking of π conjugation in the molecular backbone which induces changes in its electronic structure (energy and delocalization of molecular orbitals). In an extended π-conjugated molecule, it is also known that the conductance depends on the dihedral (torsion) angles, $\Theta$, between adjacent molecular building-block units, as it modulates the coupling between π orbitals of adjacent monomer units. Theory predicts that the conductance, G, is proportional to $\cos^2\Theta$.[17] ,[18, 19-20] This behavior has been confirmed experimentally.[21-24] In fact, these experiments are performed on a series of molecules sharing the same backbone based on a biphenyl unit. The torsion angle between the two adjacent phenyl rings is controlled by molecular engineering, through the introduction of increasingly bulky side groups, which induce a progressive increase in the dihedral angle $\Theta$ by steric hindrance. In these experiments, the conductance decreases by a factor of about 16 - 20 when $\Theta$ varies from almost 0° (planar molecule) to 90°. These experiments can be qualified as static, in the sense that the conductance switching cannot be triggered by an external stimulus (e.g. light, voltage, pH, etc).

Here, we demonstrate experimentally that the photochemical *trans-cis* isomerization of an azobenzene unit can reversibly modulate the torsion angles in an oligothiophene chain attached in two remote positions, see Fig. 1-a.[25] We show that this feature is preserved as the molecule is contacted to electrodes, allowing for a control of the torsion angle *in situ*. Consequently, our device constitutes a new type of reversible photoswitching candidate in the field of molecular electronics.

We propose the following mechanism: upon irradiation of the azobenzene at 360 nm, the isomerization from the *trans* to the *cis* configuration of azobenzene occurs and consequently, a mechanical constraint is induced in the oligothiophene chain. This leads to a conformational transition characterized by a significant change in the torsion angles along the oligothiophene chain. The latter goes together with a modification of the electronic properties of the oligothiophene unit and hence with a change in conductance of the molecule.

This mechanism is first established in solution through electrochemical and optical characterizations. In a second step, self-assembled monolayers (SAMs) of these molecules are deposited on gold surfaces and characterized by XPS, cyclic voltammetry, ellipsometry and contact angle measurements. These results assess the structural quality of the SAMs and the preservation of the isomerization reaction when the molecules are covalently bound on the metal surface. Finally, electrical characterizations are performed by contacting the monolayer with eutectic GaIn drop for two different series of junctions, following the same fabrication and measurement protocols. These devices clearly show a significant change of conductance (ratio between $10$-$10^3$) upon isomerization of the azobenzene moiety.

Ab initio calculations have been further performed to better understand the origin of the conductance switching. We identify two almost energetically similar oligothiophene conformations for the molecules with the azobenzene in the *trans* form and three for the *cis* form. The corresponding calculated conductances show that the experimental trends are reproduced by the calculations. This allows us to conclude that the change in conductance is driven by the change in the torsion angles in the quaterthiophene chain, induced *in situ* by the light-driven isomerization of the azobenzene loop.

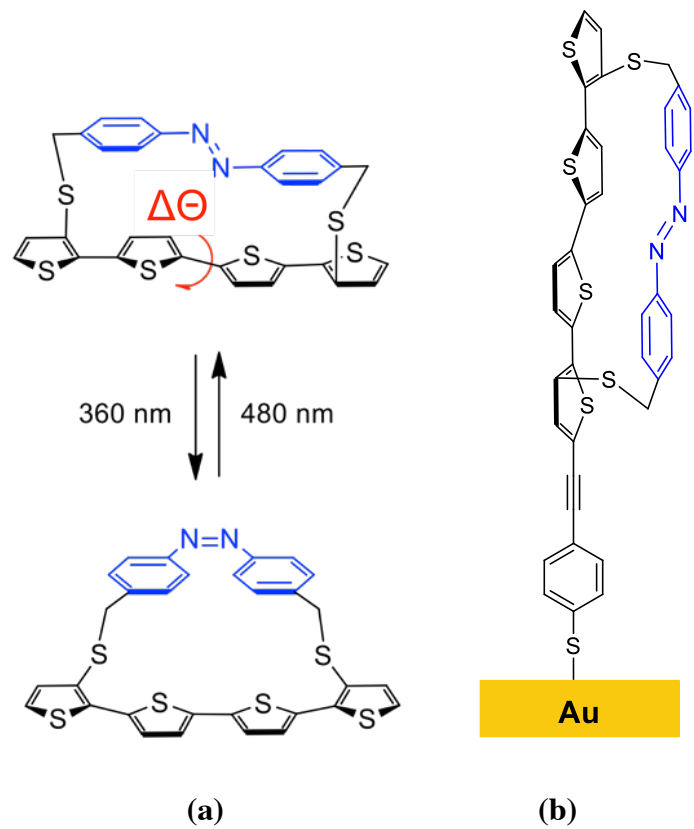

**Fig. 1**. (a) Principle of photochemically dynamic π-conjugated systems. (b) Thiol derivative designed for self-assembled monolayers on Au.

RESULTS

**Synthesis and switching properties in solution.** To fabricate the molecular junctions (MJ), the synthesis of a quaterthiophene-azobenzene derivative ("4T//azo" in the following) with a terminal thiol group is needed to graft it on a gold surface (Fig. 1-b). We have focused our attention on the synthesis of the monothiol compound **7** (see scheme 1) consisting of a quaterthiophene π-backbone with an ethyne-benzenethiol prolongation (Fig. 1-b). The synthesis of the thiol derivative **7** is described in scheme 1 and in the Method section.

The evolution of the UV-vis spectrum of compound **6**, bearing a protected thiol group, was analyzed in solution upon irradiation at 360 nm for various times (Fig. 2-a) and then at 480 nm (Fig. 2-b). The initial spectrum of compound **6** exhibits two main absorption bands with maxima at 357 and 439 nm, respectively, corresponding to the π-π* electronic transition of the azobenzene moiety and the π-π* electronic transition of the quaterthiophene-acetylene-benzene conjugated segment. Irradiation at 360 nm leads to a significant decrease of the band at 357 nm in agreement with the progressive *trans* to *cis* isomerization of the azobenzene, which is also accompanied by a slight increase of the band at 439 nm. The back *cis* to *trans* reaction is observed upon irradiation at 480 nm of the previous solution. At this wavelength, the lowest π -π* electronic transition of the azobenzene of weak intensity is involved in the back reaction although it is masked by the broad π-π* band of the quaterthiophene-acetylene-benzene backbone (Fig. 2-b).

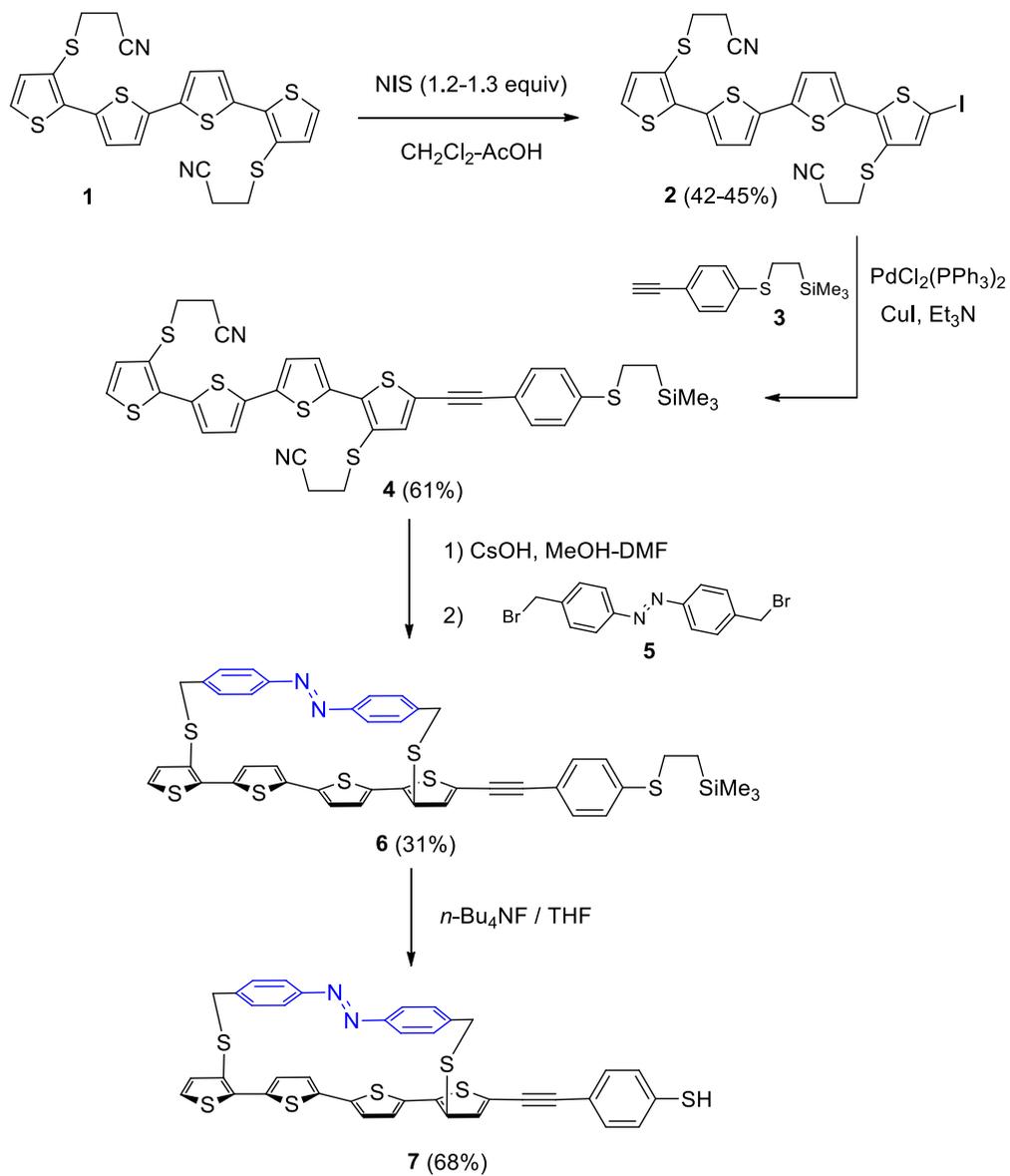

**Scheme 1.** Synthesis of monothiol **7**.

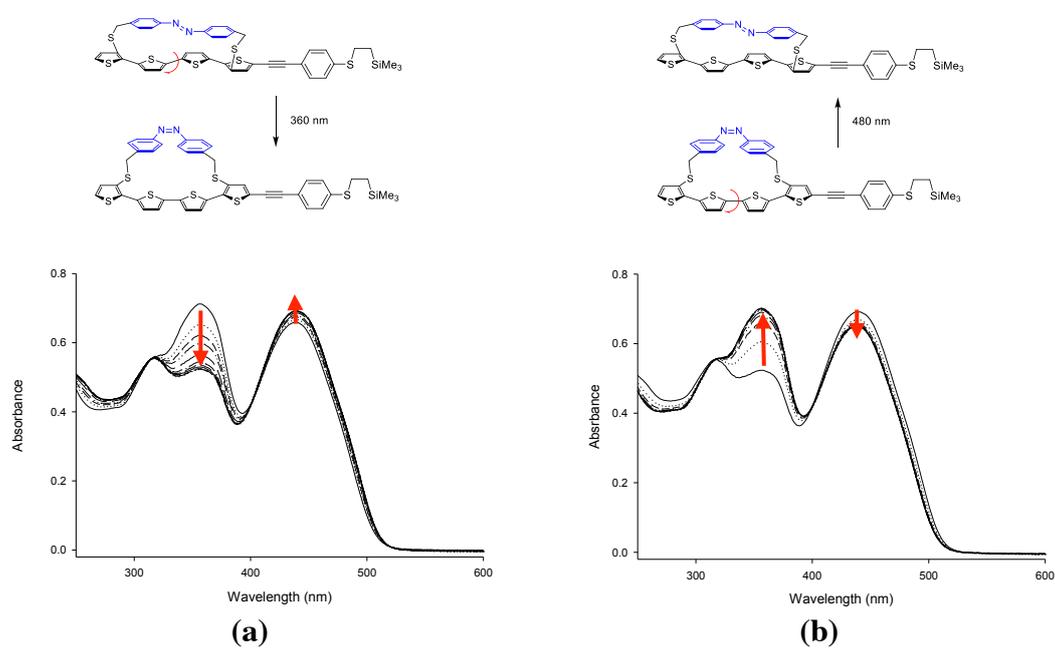

**Fig. 2.** (a) UV-vis spectra of **6** in CH$_2$Cl$_2$ (1.5 x 10$^{-5}$ M) after irradiation at 360 nm at t$_0$, +5 min, +5 min, +5 min, +10 min (repeated 7 times); (b) and irradiation at 480 nm at t$_0$, +1 min (repeated 8 times).

The reversible photoisomerization of **6** was also observed by cyclic voltammetry (Fig. 3). The initial cyclic voltammogram (CV) of a solution of **6** exhibits mainly a quasi-reversible oxidation peak at 1.02 V/SCE. This peak is related to the oxidation of the quaterthiophene-acetylene-benzene conjugated backbone of molecule **6** in its *trans*-azobenzene state. A weak shoulder is also discernible at *ca*. 0.82 V/SCE due to the electrochemical oxidation of a small amount of **6** in its *cis*-azobenzene form (Fig. 3). Irradiation of a solution of **6** at 360 nm induces a slight decrease of the intensity of the oxidation peak amplitude at 1.02 V/SCE whereas the intensity of the oxidation peak amplitude at 0.82 V/SCE increases. Note that these two distinct electrochemical responses, before and after irradiation, are fully stable upon potential cyclings between 0.2 to 1.1 V with no deviation of the peaks confirming that their difference is related to the effect of the irradiation. This behavior agrees well with a transition from the *trans* to the *cis* form in agreement with previous results.[25] This process is reversible upon irradiation at 480 nm. Due to the relative high concentration of **6** used in cyclic voltammetry and the low power of the light source (see the Supporting Information) used for irradiation, the efficiency of the photoisomerization is relatively low. However, this result shows that the *trans* and *cis*-isomers exhibit different electrochemical properties suggesting, in particular, a higher energy level of the HOMO (for the *cis*) as indicated by its slightly lower oxidation potential. This difference is expected to lead to different electrical properties of monolayers based on the *trans* versus *cis* form.

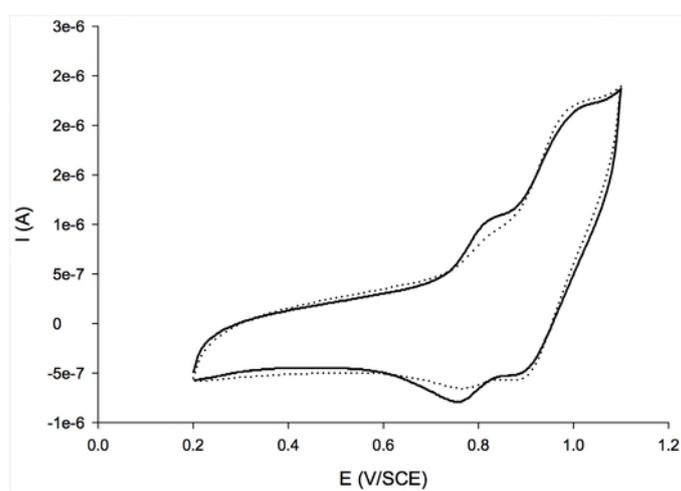

**Fig. 3**. Cyclic voltammogramms of **6** (5 x $10^{-4}$ M) in 0.10 M Bu$_4$NPF$_6$ / CH$_2$Cl$_2$,

100 mV/s, Pt electrode, before irradiation (dotted line, *trans*) and after irradiation

(solid line, *cis*) at 360 nm for 20 h.

**Self-assembled monolayers.** Self-assembled monolayers (SAMs) of thiol **7** were prepared on planar gold surfaces (see method section). The cyclic voltammetry (CV) of the monolayer shows (Fig. 4) a broad reversible oxidation wave at 1.13 V/SCE associated to the oxidation of the quaterthiophene-acetylene-benzene conjugated backbone, hence demonstrating the grafting of the molecules of **7** on the surface. Fig. 4 also presents the CV recorded at various scan rates, showing that a linear variation of the peak current *versus* scan rate is observed, thus confirming that molecules **7** are immobilized on the electrode surface. In addition, the surface coverage ($\Gamma$) of molecules **7** was determined by integration of the voltammetry peak (inset Fig. 4-b) giving a value of $\Gamma = 9.9 \pm 1 \; 10^{-11}$ mol/cm$^2$ (or $168 \pm 17$ Å$^2$ per molecule).

The thickness of the SAM in the *trans* form (e = $12.3 \pm 2$ Å) measured by ellipsometry (details in the Supporting Information) is smaller than the length of the molecule (~22 Å). From these values, we can estimate that the tilt angle of the molecule is around 55° with respect to the normal to the surface. This relatively low molecular density is in agreement with the CV measurements pointing also to a low degree of coverage (see above). The measured thickness is not significantly modified upon light exposure induced *trans-cis* isomerization.

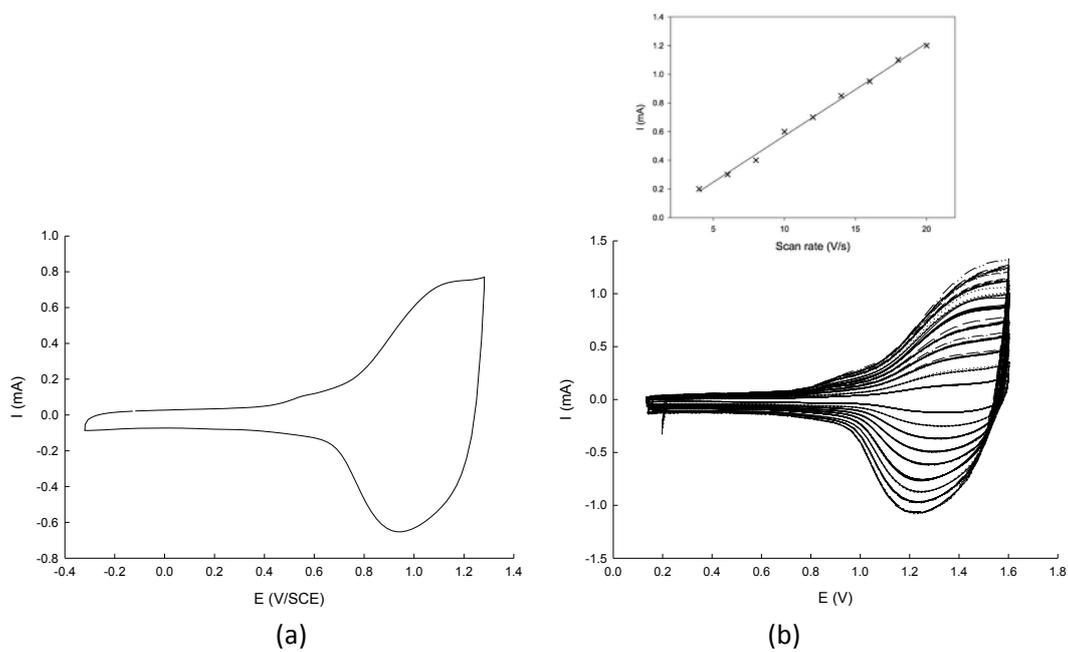

(a)                        (b)

**Fig. 4**. (a) Cyclic voltammogramms of a monolayer of **7** in 0.1 M

Bu$_4$NPF$_6$/CH$_3$CN, SCE reference electrode, 14 V s$^{-1}$ and (b) at various scan rates:

4, 6, 8, 10, 12, 14, 16, 18 and 20 V s$^{-1}$, Ag wire as pseudo reference. Inset : Linear evolution of the intensity of the CV peak (at 1.5V/SCE) vs. the scan rate.

The analysis of the XPS spectrum (details in the Supporting Information) shows only the different atoms of the molecule: Carbon (C1s), Nitrogen (N1s), Sulfur (S2s, S2p). From the integral of these peaks, the experimental atomic ratios are very close to the theoretical values considering the chemical constitution of 4T//azo (C1s/S2p=4.8±0.4 and C1s/S2s=4.5±0.4 compared to the expected ratio 38/7=5.4, C1s/N1s=21±3 compared to the expected ratio 38/2=19). From the high-resolution XPS spectrum for the S2p region (Fig. 5), we estimate the ratio of sulfur atoms linked with carbon atoms to the sulfur atoms linked with gold atoms to S2p-C/S2p-Au=6.6±1, a value consistent with the expected ratio 6/1=6 further indicating that the integrity of the molecules is preserved upon adsorption on gold.

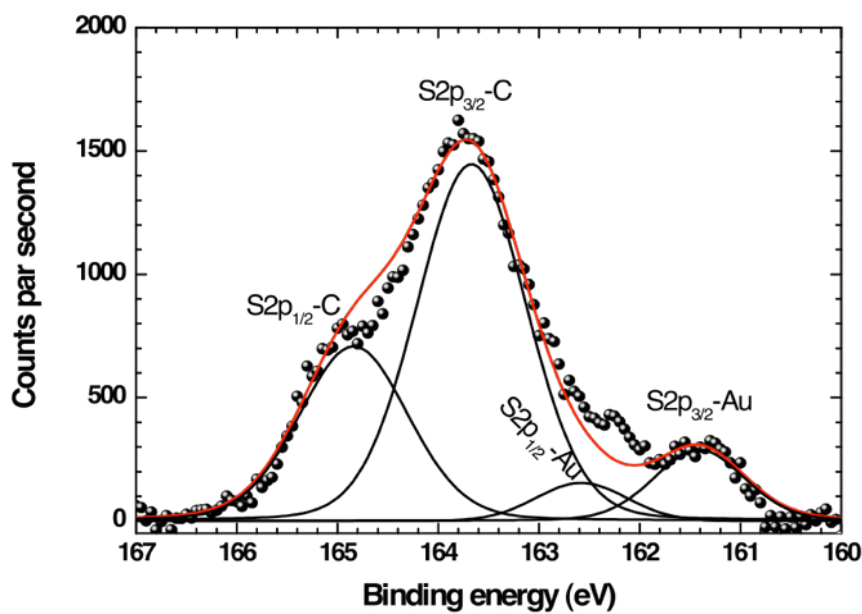

**Figure 5:** S2p region of the high-resolution XPS spectrum of monolayer of molecule **7** on Au: circles correspond to the experimental spectrum, black line to the 4 fitted peaks associated to the two doublets S2p$_{3/2}$ and S2p$_{1/2}$ for sulfur atoms linked with gold atoms and carbon atoms; the red line is the sum of the 4 peaks.

To follow the isomerization reaction in the monolayer, we measured the evolution of the water contact angle (details in the Supporting Information) of the SAM as a function of the light irradiation wavelength (Fig. 6). Before light irradiation, the contact angle on the SAM is 69 ± 1°. After UV irradiation (*trans-to-cis* isomerization) during 60 minutes (light irradiation details in the Supporting Information), the contact angle decreases to 65 ± 1°, and switches back to a higher value (70 ± 2°) after blue irradiation (*cis-to-trans* isomerization). This behavior is reproducible during two cycles with a relative stability of the contact angle value for each isomer. This variation of the contact angle between the two isomers, ΔΨ between 4-7°, with no overlapping of the error bars, is close to the one usually measured for other azobenzene systems.[5, 26] The difference of wettability between the two isomers is attributed to differences in the molecular dipole: the dipole moment of the *cis* isomer is slightly larger than for the *trans* isomer.[5,26] This behavior observed upon light illumination shows that the reversible isomerization of the molecules is preserved when they are grafted on the gold surface. This result also permits to link the transport properties (see below) with the state of the molecules in the SAMs (*cis* after UV light, and *trans* after blue light). Note that other techniques, which are classically used for this identification purpose, were not helpful in this case. Since, the length of the molecule does not change significantly upon isomerization (ca. 2.2 nm) we did not observe any change in the SAM thickness by ellipsometry measurements. CV measurements directly on the SAM were not conclusive because the oxidation peak (Fig. 4) is broad and we cannot observe the small change as in solution (Fig. 3). We also note that the two states are thermally stable for several hours at room temperature, which is sufficient to carry out the current-voltage measurements reported hereafter.

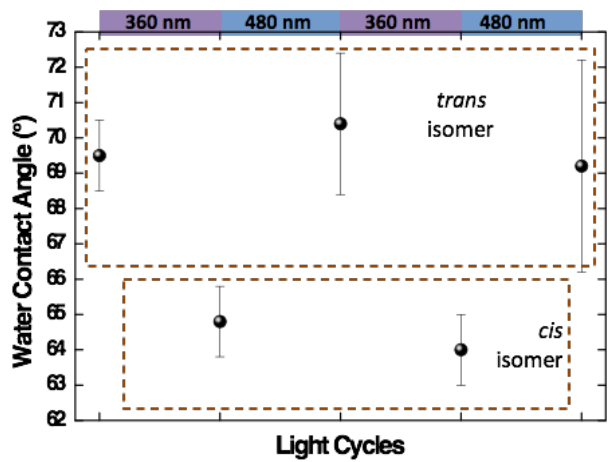

**Figure 6**. Evolution of the water contact angle of the SAM as a function of the irradiation during 2 cycles. After formation, the SAM was alternatively exposed to UV light (360nm) and visible light (480nm) during 60 minutes.

The electrical characterizations of these molecular switches were carried out by contacting the monolayer with eutectic eGaIn drops (see details in the Supporting Information).[27] The current-voltage curves (I-V) were recorded by applying the voltage on the eGaIn contact and following the voltage sweep sequence 0➟+1V / 0➟-1V repeated several times. Fig. 7-a shows typical I-V curves recorded for the pristine monolayer (after fabrication, in dark), then after irradiation with a UV light (360 nm, 1h, in purple) to induce a *trans-to-cis* isomerization of the azobenzene loop and after blue light irradiation (480 nm, 1h, in blue) to induce a *cis-to-trans* isomerization. Each panel in Fig. 7 corresponds to two different series of junctions (out of three) fabricated according to the same process and measured with the same protocols, at several month intervals. Each I-V trace corresponds to a measurement taken at a different place on the monolayer (see details in the supporting information). We first report the results of two sets of samples showing very marked differences in their electron transport measurements, the third one (showing more fuzzy results at a first glance) is discussed at the end (discussion section). In one case (Fig. 7-a), the conductance in the *trans* conformation is higher by a factor of about 3-10 than in the *cis* conformation ($G_{trans} > G_{cis}$). However, in the second set of experiments (Fig. 7-b), we observed the reverse situation, $G_{cis} > G_{trans}$, with a ratio of about $10\text{-}10^3$. Note also that the I-Vs for the pristine monolayers differ significantly between the two sets of measurements. The conductance of the pristine SAM is lower in the case $G_{cis} > G_{trans}$, with the lowest value observed in these experiments.

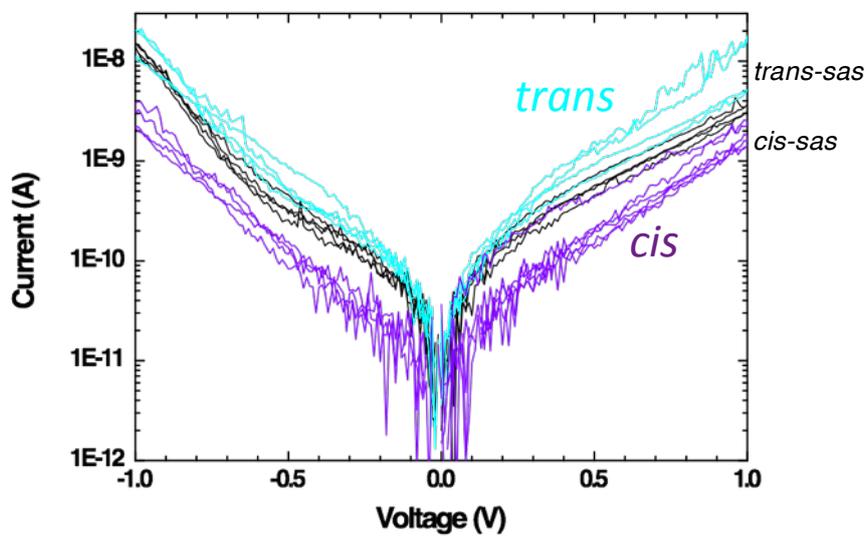

(a) $G_{trans} > G_{cis}$

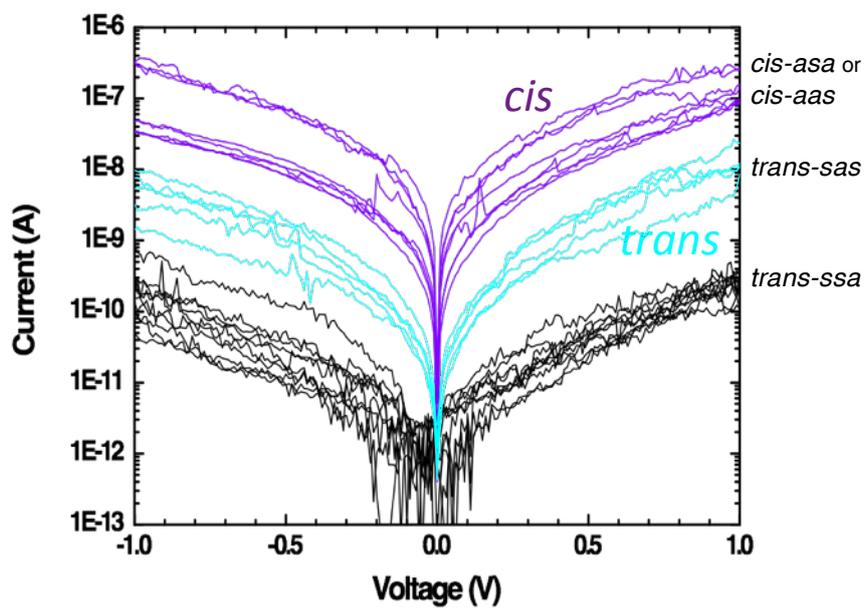

(b) $G_{cis} > G_{trans}$

**Figure 7.** (a) Typical I-V curves obtained in different regions of the sample: (dark line) before light exposure; (purple line) after UV light irradiation during 1h to switch to the cis isomer; (blue line) after blue light irradiation during 1h to switch back to the trans isomer. (b) I-V curves for a second series of molecular junctions. (dark line) before light exposure; (purple line) after UV light irradiation during 1h, isomer cis; (blue line) after blue light irradiation during 1h, isomer trans. For the isomer identification (on the right side), see the discussion section.

**Theory.** To understand the origin of these two opposite effects of isomerization on the conductance properties, the electronic properties of the 4T//azo molecule were studied theoretically with *ab initio* calculations (see Methods). Figure 8 shows the five resulting geometrical conformations for which a local minimum energy was reached by the optimization algorithm. Among these, we have identified two almost energetically similar quaterthiophene conformations for the molecules with the azobenzene moiety in the *trans* form and three for the *cis* form. They differ by the relative *syn* (s) or *anti* (a) orientation of two consecutive thiophene rings in the 4T unit : *trans-sas*, *trans-ssa*, *cis-asa*, *cis-sas* and *cis-aas*, respectiv*ely*. For the *trans* isomer, the two conformations are *trans-sas* and *trans-ssa*. The most stable conformation is *trans-sas*, which is considered as the reference in the following. The other *trans* configuration (trans-ssa) exhibits a slight increase in total energy, about 207 meV. A much larger increase is observed for the cis conformations : 716 meV for *cis-aas*, 754 meV for *cis-sas* and 759 meV for *cis-asa*.

Looking at these conformations, we see that large conformational changes are expected in the 4T backbone during a transition between the *trans-* and *cis-*azobenzene states. More specifically, the dihedral angles between two adjacent thiophene units are largely changed upon isomerization. Table 1 summarizes the calculated values for the three angles in the 4T part of the molecules. On the other hand, the impact on the molecular size is much more limited, as confirmed by the fact that the SAM thickness revealed by ellipsometry is not modified upon light irradiation.

We now analyze the conductance properties of the various conformations at the theoretical level. Qualitatively, we can describe the 4T backbone as made of four fragments, sharing the same energy, $\varepsilon$, that are coupled by an electronic coupling term, $t_i$, proportional to the cosine of the dihedral angle between two adjacent thiophene units. Since the molecules are long, we neglect here any through-space contribution to the current across the junction. Assuming that an off-resonant transport regime is operative, which is usually the case for long conjugated fragments, our description leads to a transmission obeying the following relationship:[28]

$$T(E) \approx \frac{K}{(E-\varepsilon)^8} \prod_{i=1}^{3} \cos^2 \theta_i$$

(1)

i.e., the total transmission is proportional to the product of the squares of the cosine of the torsion angles, $\theta_i$, given in Table 1. K is a constant of proportionality that represents the coupling of the external fragments with the electrodes. We can reasonably assume that the thiol contacts are equivalent for the five conformations owing to the fact that the tilt angle is constant for all molecules. From equation (1), we identify three parameters that can modify the conductance at the Fermi level among the conformations: the Fermi level alignment, the quality of the eGaIn drop contact (factor K) and the torsion angles.

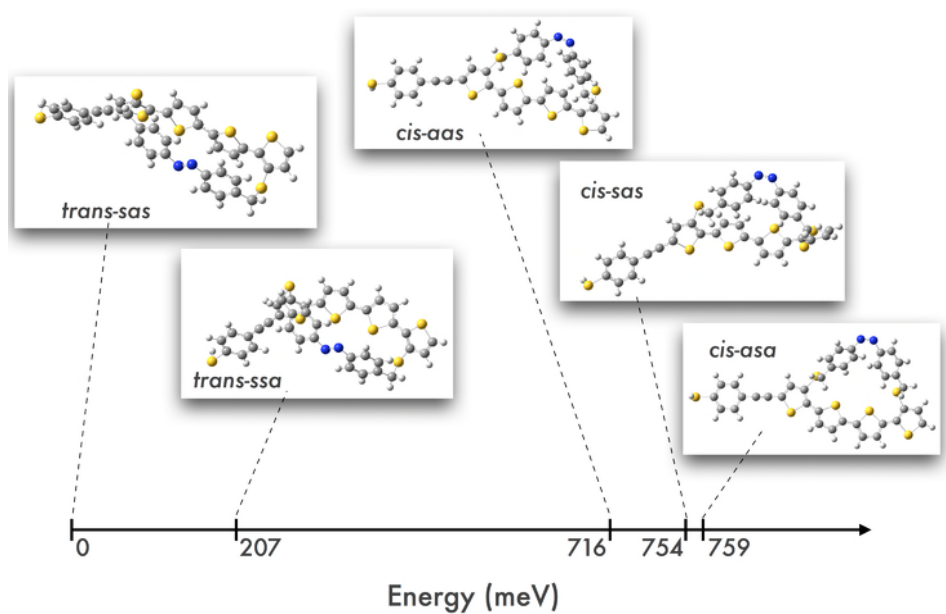

**Figure 8**: Geometrical conformations of the 4T//azo molecule for the trans and cis isomers of the azobenzene unit. The energy scale refers to the total energy considering the most stable configuration (*trans-sas*) as the reference.

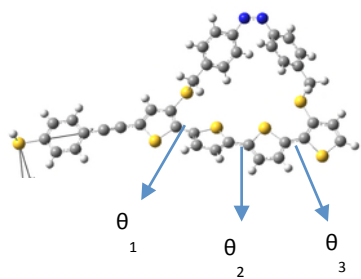

| | θ₁ (°) | θ₂(°) | θ₃ (°) | L (nm) | $\cos^2\theta_1$ | $\cos^2\theta_2$ | $\cos^2\theta_3$ | $\Pi_i \cos^2\theta_i$ |
|---|---|---|---|---|---|---|---|---|
| **cis-asa** | 23.8 | 18.8 | 7.2 | 2.42 | 0.84 | 0.90 | 0.98 | 0.74 |
| **trans-sas** | 27.8 | 10.2 | 28.7 | 2.32 | 0.78 | 0.97 | 0.77 | 0.58 |
| **trans-ssa** | 29.1 | 26.1 | 30.0 | 2.24 | 0.76 | 0.81 | 0.75 | 0.46 |
| **cis-aas** | 13.4 | 20.2 | 48.7 | 2.36 | 0.95 | 0.88 | 0.44 | 0.36 |
| **cis-sas** | 57.9 | 9.5 | 44.3 | 2.38 | 0.28 | 0.97 | 0.51 | 0.14 |

**Table 1.** Dihedral angles in the 4T moiety (θᵢ as defined in the inset), total length of the molecules (between the two terminal hydrogen atoms), corresponding $\cos^2\theta_i$ and their product. The molecules are resorted (from top to bottom) by decreasing conjugation along the backbone, i.e., decreasing product of $\cos^2\Theta_i$.

Looking at the different conformations and the last column in Table 1, we deduce that the maximal impact on the conductance that can be expected from the sole effect of the torsion angles is about 5, according to equation (1) (considering K and E-ε as constant). This is in agreement with the experimental conductance modulation, about a factor of 2, going from 0° to 48° for the torsion angle in a biphenyl molecule.[21] This implies that the modification of the torsion angles has an important impact on the molecular conductance. If this can well rationalize the ratio of conductance in the first plot of Figure 7, it cannot solely be at the origin of the experimentally observed conductance switching for the second series of junctions.

To further quantify the other parameters in equation (1), we carried out NEGF-DFT calculations of the transmission spectra for monolayers built from each conformation. In our simulations, the phenyl plane is tilted by an angle of 55°, in agreement with our SAM characterization and previous DFT theoretical calculations.[29] We used a surface area per molecule of 199 Å², a value close to the packing density suggested by the surface CV experiment. All molecules are contacted to the bottom electrode with the sulfur atom lying on-top of a gold atom of the (111) surface. Since the structure of the eGaIn drop is ill defined, we use in our model system a flat gold layer as top electrode, positioned at 2.86 Å away from the upmost hydrogen atom in the junction. Since the eGaIn contact is a weak van der Waals contact,[27, 30-31] the flat gold metal contact, at such distance from the molecules, pretty mimics the actual eGaIn contact. More computational details are provided as Supporting Information. The calculated transmission spectra at zero bias are reported in Figure 9.

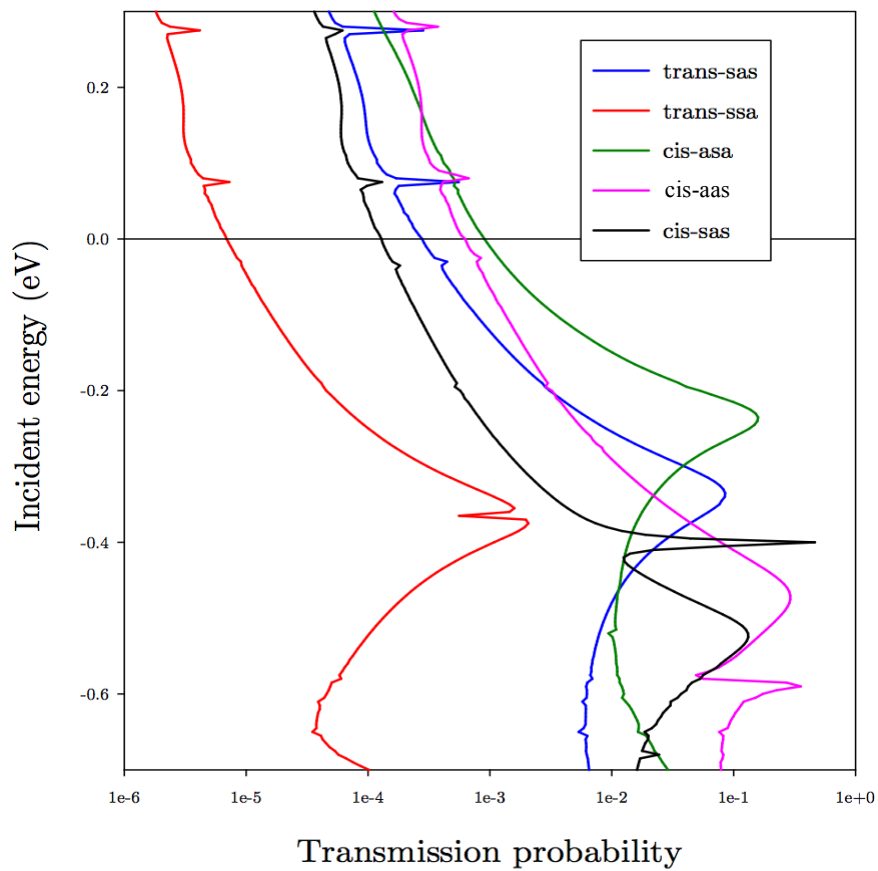

**Figure 9**. Calculated transmission probability for the 5 conformations of the 4T//azo molecule. The electrodes Fermi level is set as the zero of energy.

DISCUSSION

We observe that the spectra are dominated by the tail of the HOMO level lying below the Fermi level. The transmission probabilities, at the Fermi level, follow the order *cis-asa > cis-aas > trans-sas > cis-sas > trans-ssa*. This translates into the following zero-bias conductances using the Landauer-Buttiker expression: $G_{cis-asa}$ (= $9\times10^{-4}$ $G_0$) > $G_{cis-aas}$ (= $6.3\times10^{-4}$ $G_0$) > $G_{trans-sas}$ (= $2.8\times10^{-4}$ $G_0$) > $G_{cis-sas}$ (= $1.3\times10^{-4}$ $G_0$) > $G_{trans-ssa}$ (= $2.5\times10^{-6}$ $G_0$). If we *focus* only on the cis isomers, the order of the conductances is in agreement with the square cosine factor, Fig. 10. For the *cis* isomers, the variations are within a decade, in agreement with previously reported results for series of π-conjugated molecules (biphenyl) with "static" torsion angles.[21-22] The variation is not strictly linear as the HOMO position, E-ε in Eq. 1, is slightly varying as shown in Fig. 9. A discussion in terms of the only torsion parameters does not hold if we include the *trans* isomers in the comparison. Indeed, on the basis of the torsion angle, the two trans isomers should have a larger conductance than the cis-sas and cis-aas isomers.

This different ordering cannot be explained by the Fermi level alignments since the HOMO levels are only slightly dispersed between -0.25 and -0.5 eV below the Fermi energy. Despite their lower predicted conductance, the *trans* isomers not only exhibit a higher degree of conjugation throughout the chain (table 1) but also align closer to the Fermi level than the *cis-aas* and *cis-sas* derivatives.

In the end, the coupling between the molecule and the top contact (K factor in Eq. 1) appears to also govern the changes in the conductance upon switching between the *cis* and *trans* isomers. Especially, for the *trans* isomers, the atomic details of the contact geometry appears to be a key factor leading to a change in conductance that is larger than that predicted by Eq. (1), solely based on the modifications of the torsion angles or Fermi level alignments. All together, the switching of conductance is thus due to both the modulation of the torsion angle and the contact resistance upon isomerization.

We cannot directly compare the calculated conductance for a single molecule and the corresponding experimental values for a macroscopic junction because it is known that, due to molecule-molecule interactions, the conductance of an ensemble of molecules is different than the sum of its parts,[32-33] and that, due to defects and impurities, the exact number of molecules effectively contributing to the electron transport in a macroscopic junction is much less than the total number (effective electric contact area can be $10^3$-$10^4$ lower than the geometrical contact).[34-35] Moreover, the nature of the top contact is not the same in the experiments and simulations, thus further preventing a direct comparison. Nevertheless, these calculations show, qualitatively, a trend that can explain the experimental results presented in Fig. 7. This especially holds true if we focus on the conductance ratios, rather than on the absolute predicted conductances.

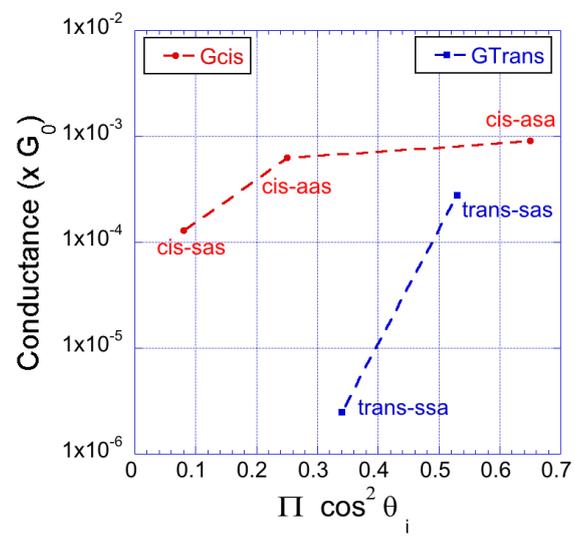

**Figure 10.** Plot of the calculated conductances versus the product of cos²Θ$_i$

Since we have 3 almost energetically equivalent *cis* conformations and 2 *trans* conformations, we can rise the hypothesis that, depending on the experiments, we can randomly observe switching between any of the 6 possible *trans/cis* combinations. The comparison between theory and experiment can then be rationalized as follows.

- The lowest calculated conductance ($2.5 \times 10^{-6}$ $G_0$ for *trans-ssa*) can be ascribed to the lowest current measured in the 2 series of experiments, i.e. for the as-fabricated SAM in Fig. 7-b (dark line).

- The highest calculated conductances ($9 \times 10^{-4}$ $G_0$ for *cis-asa* and $6.3 \times 10^{-4}$ $G_0$ for *cis-aas*) can be associated with the highest measured current after UV irradiation (*cis* isomer) in Fig. 7-b (purple line). We do not distinguish here the *cis-asa* and *cis-aas* since they have rather similar conductance (within a factor smaller than the experimental dispersion).

- Consequently, the third calculated conductance for *cis-sas* ($1.3 \times 10^{-4}$ $G_0$) would correspond to the measured current shown in Fig. 7-a after UV light irradiation (purple line), which is the last remaining experimental measurement for the *cis* isomer.

- Finally, the calculated conductance of $2.8 \times 10^{-4}$ $G_0$ for *trans-sas* can be ascribed to the other measured current in the *trans* form, i.e. before irradiation (dark line, Fig. 7-a) and after blue light irradiation (blue lines in Figs. 7-a and 7-b); in these two cases, the measured current is almost the same (around $10^{-8}$ A at +/- 1V).

Table 2 summarizes the switching ratios between the different scenarii and confirms the good qualitative agreement between experiment and theory.

Results in Figs. 7 call for several comments. For example, albeit the *cis* isomers have relatively close total energies (see Fig. 8) we have observed the same isomer at different locations on the SAM (e.g. *cis-sas* in Fig. 7-a) and not a mix of the different possible conformers. The switching behaviors are also occurring between specific conformers (*trans-sas → cis-sas → trans-sas* for sample Fig. 7-a, and *trans-ssa → cis-asa* (or *cis-aas* or a mix of the two) → *trans-sas* for sample in Fig. 7-b). However, we have also observed a "mixed" situation for a third set of samples/measurements (Fig. S8 in the supporting information). In that case, the I-V traces seem fuzzier at first glance, but based on the theoretical results and the identification of specific conformers for the results showed in Fig. 7, we can conclude that each *trans* and *cis* conformers are observed at different locations on the SAMs upon the *trans-cis* isomerization. In that case, this result implies that we have a mixed SAM with phase separation between the different conformers instead of a homogeneous mix (see Fig. S8-b, with two zones ascribed to *trans-sas* and one to *trans-ssa*). This is a reasonable hypothesis because it is known that binary SAMs usually form phase-separated structures.[36] Moreover, it has also been observed that the composition of the binary SAM does not reflect the one in solution and that the SAM can be constituted mainly with one compound even with a 50/50 composition in the solvent.[37] Thus, we can reasonably postulate that small changes in the ambient conditions during the SAM fabrication at several month intervals (e.g. temperature, small changes of the hygrometry in the glove-box or quality of the solvents used, etc…), not controlled in our ordinary laboratory conditions, can result in several composition of the pristine SAMs of the three sample sets as described in Fig. 7-a (mainly *trans-sas*), Fig. 7-b (mainly *trans-ssa*) and Fig. S8-a (mix of both isomers). Similarly, the switching upon UV light exposure can strongly depend on the initial structure with mainly a *trans-sas* → *cis-sas* isomerization for one sample (Fig. 7-a), a *trans-ssa* → *cis-asa* (or *cis-aas*) in case of Fig. 7-b, and a mix of *cis* isomers (Fig. S8-c) when starting from a non-homogeneous structure of *trans* isomers as depicted (Fig. S8-b) for the third sample. Finally, a brief statistical analysis of the ca. 100 I-V traces recorded on the three sets of samples shows that the *trans-sas* isomer is observed in 62% (38% for the *trans-ssa*) in agreement with the lowest total energy of the former one (by ca. 200 meV, see Fig. 8). The occurrences are more balanced for the *cis* isomer (52% for the cis-sas, and 48% for the *cis-asa* and *cis-aas* (these two last ones being not distinguishable from conductance measurements). Again this is not surprising since the total energies for the three *cis* isomers are close (with a window of ca. 40 meV, Fig. 8).

CONCLUSION

In conclusion, we demonstrated the concept that the conductance of a linear π-conjugated molecule can be dynamically modified by the photochemical *trans-cis* isomerization of an azobenzene unit fixed in a lateral loop. We identified the most energetically stable conformations of the proposed molecular switch. The calculated conductances explain the trends observed from photoswitching experiments with experimental conductance switching ratios up to about $10^3$. The conductance switching is due to both the modulation in the torsion angles in the oligothiophene chain and in the contact resistance. Due to the existence of multiple configurations with almost the same local minimum energy, several switching scenarii were identified theoretically and experimentally. Strategies for further possible improvements, towards a better control of the switching behavior, include the fusion of thiophene units 2 by 2 to rigidify the molecule and limit the number of stable configurations.

| Switching ratios | Theory | Experiment |
|---|---|---|
| trans-sas/cis-sas | 2.1 | 3-10 (Fig. 7-a at +/- 1V) |
| cis-asa (or cis-aas)/trans-sas | 2.2 - 3.2* | 10-50 (Fig. 7-b at +/- 1V) |
| cis-asa (or cis-aas)/trans-ssa | 252-360* | ca. $10^3$ (Fig. 7-b at +/- 1V |

* The highest value is for the cis-aas.

**Table 2.** Switching ratios between the different cis/trans switching scenarii. Theoretical values are given from the calculated conductance while the experimental values are calculated from averaged current values at 1V and -1V, as shown in Fig. 7.

METHODS

**Chemical synthesis.** A selective monoiodination of quaterthiophene **1** [38] in the presence of *N*-iodosuccinimide leads to compound **2** in modest yields. Further Sonogashira coupling with the acetylenic compound **3** [39] gives the extended π-conjugated system **4**. Elimination of the two 2-cyanoethyl protecting groups of **4** in the presence of cesium hydroxide affords two thiolate groups which can react with one equivalent of bis-*para*-(bromomethyl)azobenzene **5** [40] in high dilution conditions, giving the product of macrocyclization **6** in 31% yield. Finally, the thiol derivative **7** group is formed after deprotection of the 2-trimethylsilylethyle group by treatment of **6** with tetrabutylammonium fluoride. More details and characterization of the compounds are given in the Supporting Information (NMR spectra, Figs. S1-S7).

**Self-assembled monolayers**. SAMs of thiol **7** on planar gold surface were prepared (see method section) by immersion of the Au surface into a solution of **7** (1 mM) in $CH_2Cl_2$ for 1-3 days under a nitrogen atmosphere (details in the Supporting Information). The SAMs were rinsed with $CH_2Cl_2$ and analyzed by cyclic voltammetry, contact angle, ellipsometry and XPS measurements (see details in the Supporting Information).

**Theory.** Starting from the experimental crystalline structure extracted from X-Ray characterization,[25] we calculated the conformations and electronic structures of the 4T//azo molecule in the *trans* and *cis* states, at the Density Functional

Theory (DFT) level, using the B3LYP functional and a 6-311G basis set (see details in the Supporting Information).

ASSOCIATED CONTENT

**Supporting Information**. Synthetic procedures and compounds characterizations, NMR spectra, self-assembled monolayer fabrication and characterization (spectroscopic ellipsometry, XPS measurements, contact-angle measurements, light exposure protocols, electrical transport with eGaIn drop contact), details on theory are available free of charge and I-V curves for the 3[rd] set of samples.

AUTHOR INFORMATION

**Corresponding Author**

* Corresponding authors : dominique.vuillaume@iemn.fr ;
stephane.lenfant@iemn.univ-lille1.fr

**Author Contributions**

S.L. and Y.V. did the physical characterizations of the SAMs and electrical measurements. C.K., C.V.D. and J.C. did the DFT and NEGF-DFT calculations. D.D., I.D., M.O did the chemical synthesis, the characterization of the compounds and the chemical characterizations of the SAMs. P.B., J.R and D.V. supervised the project. D.V. wrote the papers with the contributions of all authors.

**Notes**

The authors declare no competing financial interest.

**Acknowledgements**

We acknowledge financial support from EU FP7 FET project SYMONE (grant 318597) and ANR project SYNAPTOR (grant 12BS0301001). C.V.D. was supported by a Gustave Boël – Sofina Fellowship of the Belgian American Educational Foundation (BAEF). The work in Mons has been supported by the Belgian National Fund for Scientific Research. J.C. is an FNRS research director.

**ToC graphic.**

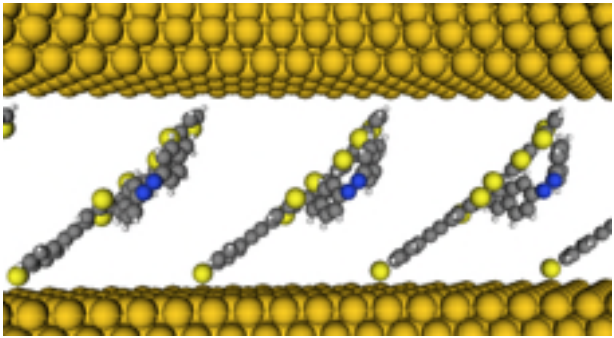

# A New Photomechanical Molecular Switch
# Based on a Linear π-Conjugated System

S. Lenfant*,  Y. Viero, C. Krzeminski, D. Vuillaume*
*Institute for Electronics Microelectronics and Nanotechnology (IEMN),*
*CNRS, Avenue Poincaré,*
*Villeneuve d'Ascq, 59650, France.*

D. Demeter, I. Dobra, M. Oçafrain, P. Blanchard, J. Roncali
*MOLTECH-Anjou, CNRS, Univ. Angers,*
*2 Bd. Lavoisier,*
*Angers, 49045, France*

C. Van Dyck
Chemistry Department – Northwestern University
2145 Sheridan Road, Evanston, Il 60208

J. Cornil
*Laboratory for Chemistry of Novel Materials, University of Mons,*
*20 Place du Parc*
*Mons, 7000, Belgium*

*\* Corresponding authors : dominique.vuillaume@iemn.fr. ;*
*stephane.lenfant@iemn.univ-lille1.fr*

## SUPPORTING INFORMATION

1. Synthetic procedures and compounds characterizations
- General information
- 5-iodo-3,3'''-Bis (2-cyanoethylsulfanyl)-2,2':5',2'':5'',2'''-quaterthiophene (2)
- 5-((4-((2-(Trimethylsilyl)ethyl)thio)phenyl)ethynyl-3,3'''-bis(2-cyanoethylsulfanyl)-2,2':5',2'':5'',2'''-quaterthiophene (4)
- Macrocycle 6
- Macrocyclic thiol 7
- 2-(Trimethylsilyl)ethyl-4'-bromophenyl sulfide (8)
- 2-(Trimethylsilyl)ethyl-4'-iodophenyl sulfide (9)
- 2-(Trimethylsilyl)ethyl-4'-[(trimethylsily)ethynyl]phenyl sulfide (10)

- 2-(Trimethylsilyl)ethyl-4'-(ethynyl)phenyl sulfide (3)

2. NMR spectra

3. Monolayer fabrication and characterization
- SAM fabrication
- Spectroscopic ellipsometry
- XPS measurements
- Contact-angle measurements
- Light exposure
- Electrical transport with eGaln drop contact

4. Theory

5. I-V curves for 3$^{rd}$ set of samples

## 1. Synthetic procedures and compounds characterizations

**General information.** All reagents from commercial sources were used without further purification. Solvents were dried and purified using standard techniques. Column chromatography was performed with analytical-grade solvents using Aldrich silica gel (technical grade, pore size 60 Å, 230-400 mesh particle size). Flexible plates ALUGRAM® Xtra SIL G UV254 from MACHEREY-NAGEL were used for TLC. Compounds were detected by UV irradiation (Bioblock Scientific) or staining with diiode, unless otherwise stated.

NMR spectra were recorded with a Bruker AVANCE DRX 500 ($^1$H, 500 MHz and $^{13}$C, 125 MHz). Chemical shifts are given in ppm relative to TMS and coupling constants J in Hz. Residual non-deuterated solvent was used as an internal standard. Infrared spectra were recorded on a Bruker spectrometer Vertex 70. UV-Vis absorption spectra were recorded at room temperature on a Perkin Elmer Lambda 950 spectrometer. Matrix

Assisted Laser Desorption/Ionization was performed on MALDI-TOF MS BIFLEX III Bruker Daltonics spectrometer using dithranol or DCTB as matrix.

Cyclic voltammetry was performed using a Biologic SP-150 potentiostat with positive feedback compensation in dichloromethane solutions purchased from Carlo Erba (HPLC grade). Tetrabutylammonium hexafluorophosphate (0.1 M as supporting electrolyte) was purchased from Sigma-Aldrich and recrystallized prior to use. Solutions were degassed by argon bubbling prior to each experiment. Experiments were carried out in a one-compartment cell equipped with platinum working microelectrode ($\emptyset$ = 2 mm) and a platinum wire counter electrode. A wire of silver in 0.1 M TBAPF$_6$/CH$_2$Cl$_2$ was used as pseudo reference and checked against the ferrocene/ferricinium couple (Fc/Fc$^+$), the values of oxidation potentials being expressed *versus* the SCE reference electrode (E°(Fc/Fc$^+$) = 0.41 V/SCE).

Irradiation was performed with a 150 W high-pressure Xenon lamp by using band pass filters. The incident light power was 0.64 mW.cm$^{-2}$ at 360 nm and 1.52 mW.cm$^{-2}$ at 480 nm.

**5-iodo-3,3'''-Bis (2-cyanoethylsulfanyl)-2,2':5',2'':5'',2'''-quaterthiophene (2)**

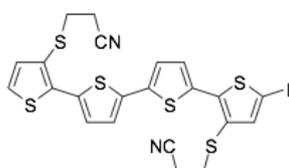

A solution of 3,3'''-bis(2-cyanoethylsulfanyl)-2,2':5',2'':5'',2'''-quaterthiophene (1 g, 2 mmol) in 150 mL of dichloromethane and glacial acetic acid (0.76 mL, 6.6 equiv) was cooled to 0 °C, and *N*-iodosuccinimide (0.584 g, 2.6 mmol) was added. The ice bath was removed, and the reaction mixture was allowed to stir at room temperature in the dark. After 21 hours, the solution was washed with a saturated aqueous solution of Na$_2$CO$_3$ (4 × 100 mL) and dried with MgSO$_4$. The solvent was removed by rotary evaporation, and the resulting solid was purified using column chromatography (silica gel/CH$_2$Cl$_2$) to provide 0.53-0.57 g (42-45%) of a yellow-orange solid. m.p. = 122-125 °C; [1]H NMR (500 MHz, CDCl$_3$) δ = 7.33 (d, 1H, $^3$J= 3.5 Hz), 7.25 (d, 1H, $^3$J= 3.5 Hz), 7.24 (d, 1H, $^3$J= 5.5 Hz), 7.21 (s, 1H), 7.17 (d, 1H, $^3$J= 4 Hz), 7.16 (d, 1H, $^3$J= 3.5 Hz), 7.08 (d, 1H, $^3$J= 5.5 Hz), 3.04 (t, 2H, $^3$J= 7.0 Hz), 3.03 (t, 2H, $^3$J= 7.0 Hz ), 2.59 (t, 2H, $^3$J= 7.0 Hz ), 2.57 (t, 2H, $^3$J= 7.0 Hz ); [13]C NMR (125 MHz, CDCl$_3$) δ = 145.48, 142.43, 139.60, 138.61, 137.70, 134.24, 133.66, 132.66, 128.21, 127.84, 125.19, 124.18, 124.07, 123.96, 118.08, 117.94, 71.92 (C-I), 32.02, 31.85, 18.68, 18.63; MALDI-TOF: m/z 626.1 [M$^+$]; IR (neat): = 2248 cm$^1$ (CN); UV-Vis (CH$_2$Cl$_2$): λ$_{max}$ = 410 nm.

## 5-((4-((2-(Trimethylsilyl)ethyl)thio)phenyl)ethynyl-3,3'''-bis(2-cyanoethylsulfanyl)-2,2':5',2'':5'',2'''-quaterthiophene (4)

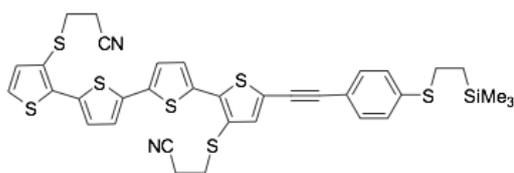

Compound **3** (0.185 g, 0.790 mmol) was added to a solution of **2** (0.45 g, 0.719 mmol) in Et$_3$N/THF (2:1) (70 mL). This solution was degassed and stirred under N$_2$. Bis(triphenylphosphine)palladium(II) dichloride (25 mg, 0.036 mmol, 5% mol.) and copper iodide (13.7 mg, 0.072 mmol, 10% mol.) were added, and the reaction was allowed to stir over night at rt. The solvent was removed by rotary evaporation, and the residue was purified by column chromatography (silica gel/CH$_2$Cl$_2$) to provide 0.32 g (61%) of a yellow-orange solid. m.p.

= 115 °C; $^1$H NMR (500 MHz, CDCl$_3$) δ = 7.41 (d, 2H, $^3$J= 8 Hz), 7.33 (d, 1H, $^3$J= 4Hz), 7.32 (d, 1H, $^3$J= 4 Hz), 7.24 (d, 2H, $^3$J= 8 Hz), 7.24 (d, 1H, $^3$J= 5 Hz), 7.20 (s, 1H), 7.18 (d, 1H, $^3$J= 4 Hz), 7.16 (d, 1H, $^3$J= 4Hz), 7.08 (d, 1H, $^3$J= 5 Hz), 3.06 (t, 2H, $^3$J= 7.0 Hz), 3.04 (t, 2H, $^3$J= 7.0 Hz), 3.01-2.97 (m, 2H), 2.61 (t, 2H, $^3$J= 7.0 Hz), 2.57 (t, 2H, $^3$J= 7.0 Hz), 0.96-0.93 (m, 2H), 0.06 (s, 9H); $^{13}$C NMR (125 MHz, CDCl$_3$) δ = 140.56, 139.57, 138.60, 137.76, 137.59, 134.20, 133.63, 133.27, 131.82, 128.18, 127.82, 127.63, 124.14, 124.03, 123.99, 123.79, 121.57, 118.93, 118.08, 117.97, 95.68, 81.77, 31.83, 28.82, 18.62, 18.61, 16.64, -1.63; MALDI-TOF: m/z 732.13 [M$^+$]; IR (neat):  = 2250 cm$^{-1}$ (CN); UV-Vis (CH$_2$Cl$_2$): λ$_{max}$ = 431, 300 nm.

### Macrocycle 6

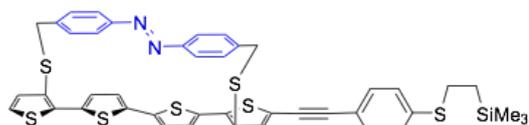

A solution of CsOH·H$_2$O (0.194 g, 1.16 mmol) in 5 mL of degassed methanol was added dropwise to a solution of **4** (0.34 g, 0.46 mmol) in 35 mL of degassed DMF. After 1 h stirring at rt, this mixture and a solution of (*E*)-1,2-bis(4-(bromomethyl)phenyl)diazene **5** (0.17 g, 0.46 mmol) in 40 mL of DMF were introduced simultaneously during 4 h using a perfusor pump in a round-bottomed flask containing 100 mL of degassed DMF at rt. Note that compound **5** has been synthesized according to a previous reference.[1] After completion of the introduction of reagents, the mixture was stirred at RT for 24 h. The solution was diluted with CH$_2$Cl$_2$, washed with water, dried over Na$_2$SO$_4$, and concentrated. The residue was purified by column chromatography (silica gel/PE/CH$_2$Cl$_2$ (1:1)) to give 0.12 g (31%) of a yellow-orange solid. m.p. = 122 °C; $^1$H NMR (500 MHz, CDCl$_3$) δ = 7.52-7.49 (m, 4H), 7.45 (d, 2H,

$^3$J= 8.5 Hz), 7.34 (s, 1H), 7.27 (d, 2H, $^3$J= 8.5 Hz), 7.235 (d, 1H, $^3$J= 4 Hz), 7.233 (d, 1H,
$^3$J= 5.5 Hz ), 7.199 (d, 1H, $^3$J= 5.5 Hz), 7.196 (d, 1H, $^3$J= 4Hz), 6.82 (br. m, 4H), 6.57 (d, 1H, $^3$J=
4 Hz), 6.55 (d, 1H, $^3$J= 4 Hz), 3.81 (s, 2H), 3.78 (s, 2H), 3.02-2.99 (m, 2H), 0.98-0.95 (m,
2H), 0.08 (s, 9H); $^{13}$C NMR (125 MHz, CDCl$_3$) δ = 151.48, 151.43, 143.46, 142.11, 141.08,
140.83, 140.78, 139.33, 136.48, 136.43, 135.55, 134.12, 133.01, 131.83, 129.03, 127.78,
127.40, 126.78, 124.20, 124.01, 123.60, 123.34, 120.94, 119.29, 95.20, 82.23, 43.47,
43.31, 28.93, 16.73, -1.60; MALDI-TOF: m/z 833.15 [M+H$^+$]; HRMS (FAB): m/z 832.0676,
exact mass for C$_{43}$H$_{36}$N$_2$S$_7$Si: m/z 832.0693; UV-Vis (CH$_2$Cl$_2$): λ$_{max}$ = 438, 357 nm.

**Macrocyclic thiol 7**

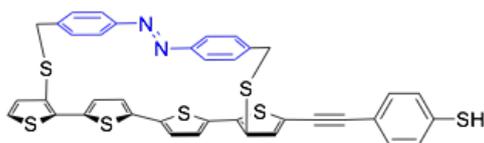

Under Ar atmosphere in an oven-dried round-bottom flask equipped with a magnetic stirrer, compound **6** (0.12 mmol, 100 mg) was dissolved in 10 mL of anhydrous THF under sonication. Then TBAF 1 M in THF (0.14 mmol, 0.14 mL) was added and the mixture was stirred under sonication for another 1.5 h (completion monitored by TLC). Upon completion, the reaction mixture was diluted with a 1 M aqueous solution of HCl (0.24 mmol, 0.24 mL), the solvent was evaporated and the residue was diluted with CH$_2$Cl$_2$ (50 mL). The resulting solution was washed with water and the organic layer was separated and then dried over MgSO$_4$. The solvent was evaporated and the residue was chromatographed (silica gel/CH$_2$Cl$_2$/PE (1:1)) affording monothiol **7** as a yellow solid (60 mg, 68% yield). $^1$H NMR (500 MHz, CDCl$_3$) δ = 7.45-7.32 (m, 7H), 7.27-7.22 (m), 7.19 (d, 2H, $^3$J= 8Hz), 6.81 (br. m, 4H), 6.59 (d, 1H, $^3$J= 4 Hz), 6.57 (d, 1H, $^3J$= 4 Hz), 3.81 (s, 2H), 3.79 (s, 2H); MALDI-TOF: m/z 732.0 [M$^+$]; UV-Vis (CH$_2$Cl$_2$): $\lambda_{max}$ = 440, 355 nm.

Compound **3** has been prepared according to a known procedure[2] as described in Scheme S1.

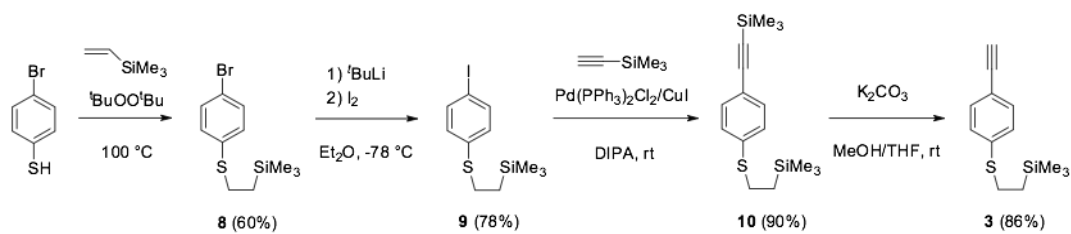

***Scheme S1.*** *Synthesis of compound **3**.*

### 2-(Trimethylsilyl)ethyl-4′-bromophenyl sulfide (8)

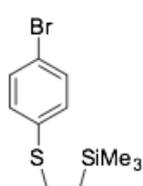

In a Schlenk tube under N$_2$, a mixture of 4-bromothiophenol (13.5 g, 71.4 mmol), vinyltrimethylsilane (12.13 mL 82.3 mmol) and *tert*-butyl peroxide (1.82 mL, 9.95 mmol) was stirred at 100 °C for 10 h. The reaction mixture was diluted by adding 200 mL of hexane, and the solution was washed once with a 10% aqueous sodium hydroxide solution. The organic layer was separated, dried over MgSO$_4$, and then concentrated. The crude product was purified by column chromatography (silica gel/Hexane) affording the title compound (13.9 g, 69%) as colorless oil. $^1$H NMR (500 MHz, CDCl$_3$) δ = 7.40 (d, 2H, $^3J$= 8.5 Hz), 7.16 (d, 2H, $^3J$= 8.5 Hz), 2.94-2.91 (m, 2H), 0.92-0.89 (m, 2H), 0.03 (s, 9H).

### 2-(Trimethylsilyl)ethyl-4′-iodophenyl sulfide (9)

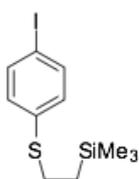

A solution of **8** (5.31 g, 18.35 mmol) in 100 mL of anhydrous diethylether was cooled to -78 °C in the presence of $N_2$. To this solution was added 18.4 mL of 1.5 M $^tBuLi$ in hexane solution, and the reaction mixture was stirred at -78 °C for 40 min. A second solution of iodine (6 g, 23.86 mmol) in 100 mL of anhydrous diethylether was cooled to -78 °C under $N_2$, and was then transferred into the cooled reaction mixture of **8**. After stirring for 30 min at -78 °C the reaction mixture was warmed to 0 °C and stirring was continued for another 30 min. The reaction mixture was then washed with a sodium thiosulfate solution until the iodine color disappeared. The diethylether layer was separated, dried over sodium sulfate and concentrated. The crude product was purified by column chromatography (silica gel column/Hexane) to give compound **9** (4.81 g, 78%) as yellow oil. $^1H$ NMR (500 MHz, $CDCl_3$) δ = 7.58 (d, 2H, $^3J$= 8.5 Hz), 7.03 (d, 2H, $^3J$= 8.5 Hz), 2.94-2.91 (m, 2H), 0.92-0.89 (m, 2H), 0.03 (s, 9H).

### 2-(Trimethylsilyl)ethyl-4′-[(trimethylsily)ethynyl]phenyl sulfide (10)

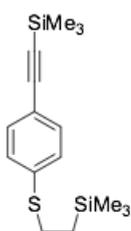

Trimethylsilylacetylene (1.54 g, 2.24 mL, 15.73 mmol) was added to a solution of **9** (4.81 g, 14.3 mmol) in diisopropylamine (200 mL). This solution was stirred under $N_2$ for 30 min. Bis(triphenylphosphine)palladium(II) dichloride (0.502 g, 0.715 mmol) and copper iodide (0.272 g, 1.43 mmol) were added, and the reaction was stirred at 50 °C for 1.5 h. The reaction was quenched with water, the aqueous layer was extracted with diethylether, and the organic extracts were washed with brine. The ether layers were dried over magnesium sulfate. The solvent was removed by rotary evaporation, and the residue was purified by column chromatography (silica gel/1:2 $CH_2Cl_2$/hexane) to provide compound **10** (3.93 g, 90%) as yellow oil. [1]H NMR (500 MHz, CDCl$_3$) δ = 7.36 (d, 2H, [3]J= 8.5 Hz), 7.18 (d, 2H, [3]J= 8.5 Hz), 2.97-2.93 (m, 2H), 0.93-0.89 (m, 2H), 0.23 (s, 9H), 0.04 (s, 9H).

### 2-(Trimethylsilyl)ethyl-4′-(ethynyl)phenyl sulfide (3)

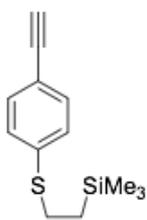

Potassium carbonate (2.85 g, 20.71 mmol) was added to a solution of **10** (3.92 g, 12.78 mmol) in MeOH/THF (1:1) (100 mL). The resulting suspension was stirred at rt for 2 h and diluted by adding 200 mL of CH$_2$Cl$_2$ and 200 mL of water. The organic layers were separated, dried over magnesium sulfate and concentrated. The crude product was purified on a short silica gel column using hexane as eluent. The desired fractions were concentrated to afford compound **3** (2.49 g, 86%) as yellow-orange oil. [1]H NMR (500 MHz, CDCl$_3$) δ = 7.39 (d, 2H, [3]J= 8 Hz), 7.21 (d, 2H, [3]J= 8 Hz), 3.07 (s, 1H), 2.98-2.94 (m, 2H), 0.94-0.91 (m, 2H), 0.04 (s, 9H).

## 2. NMR spectra

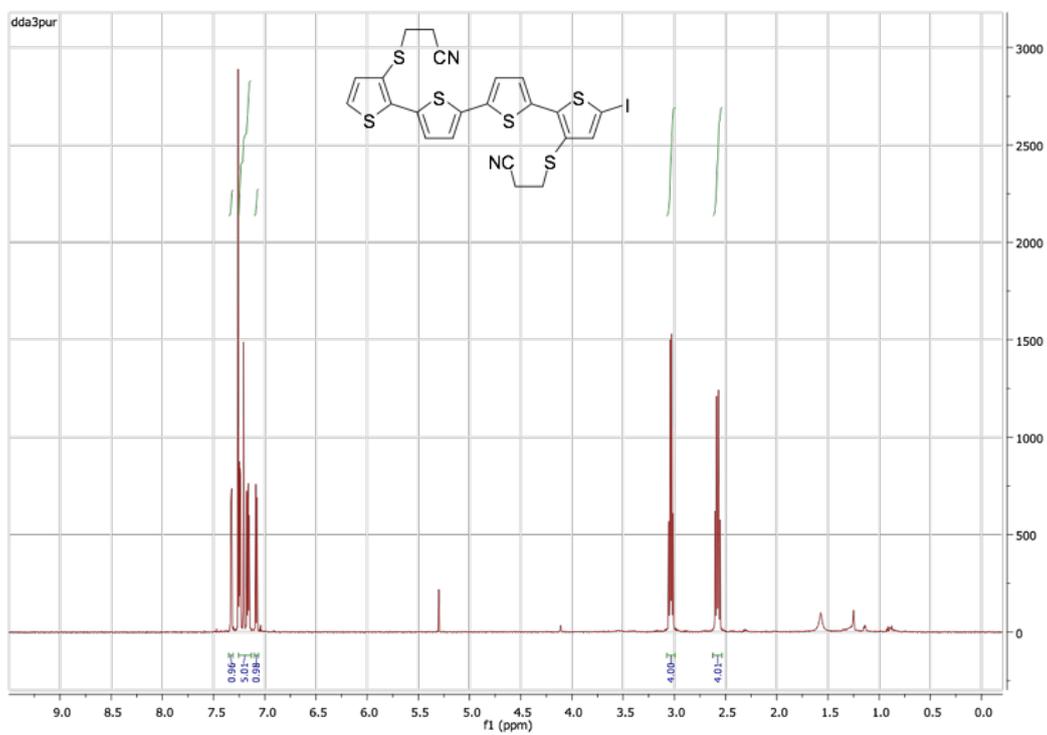

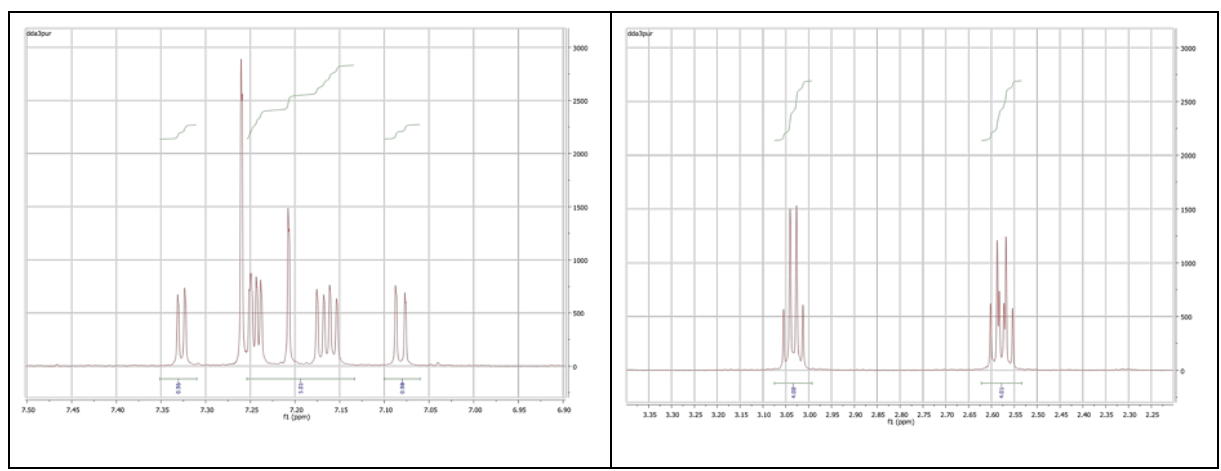

**Fig. S1.** *500 MHz $^1$H NMR spectrum of compound **2** in CDCl$_3$ at 20°C.*

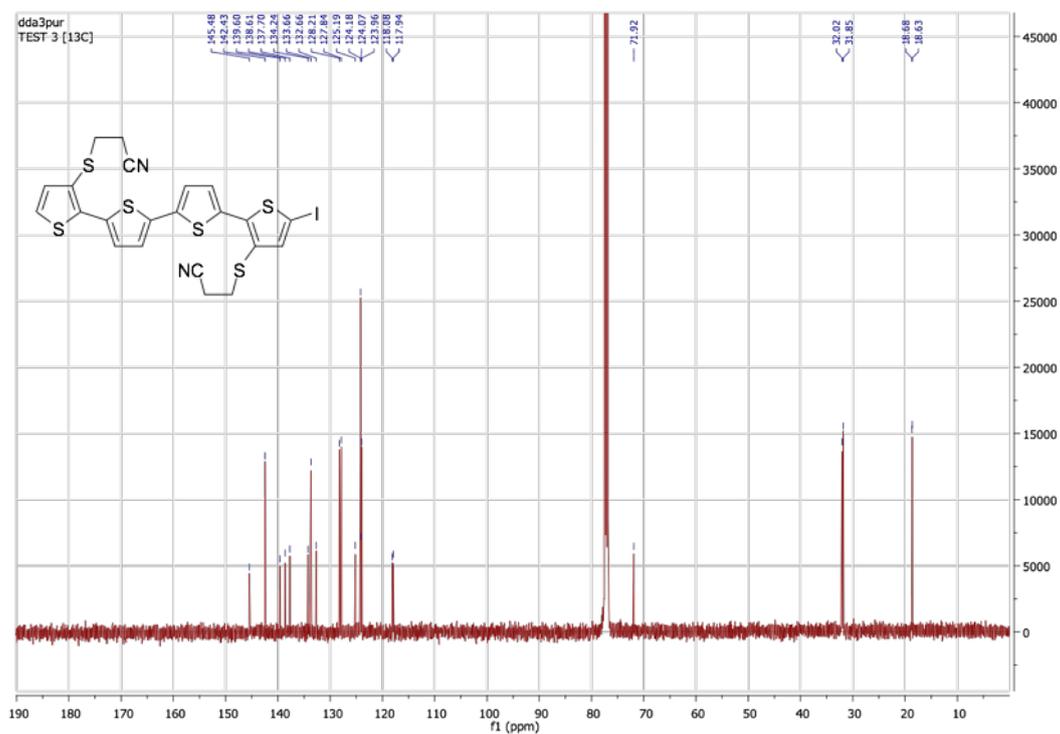

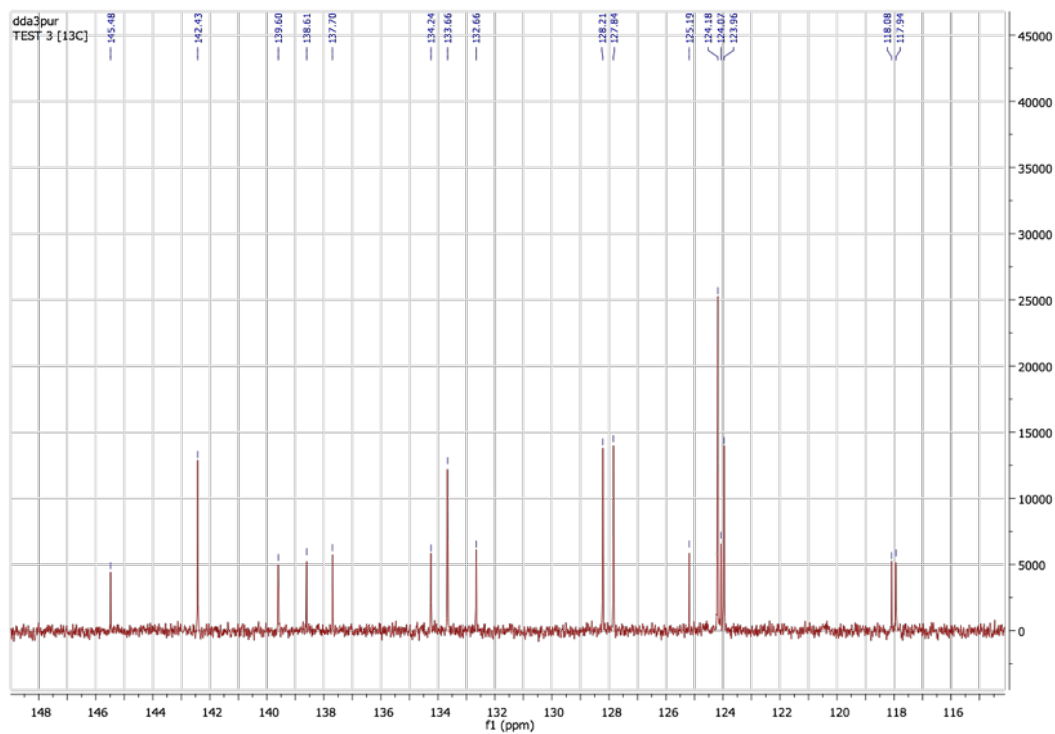

**Fig. S2.** 125 MHz $^{13}C$ NMR spectrum of compound **2** in CDCl$_3$ at 20°C.

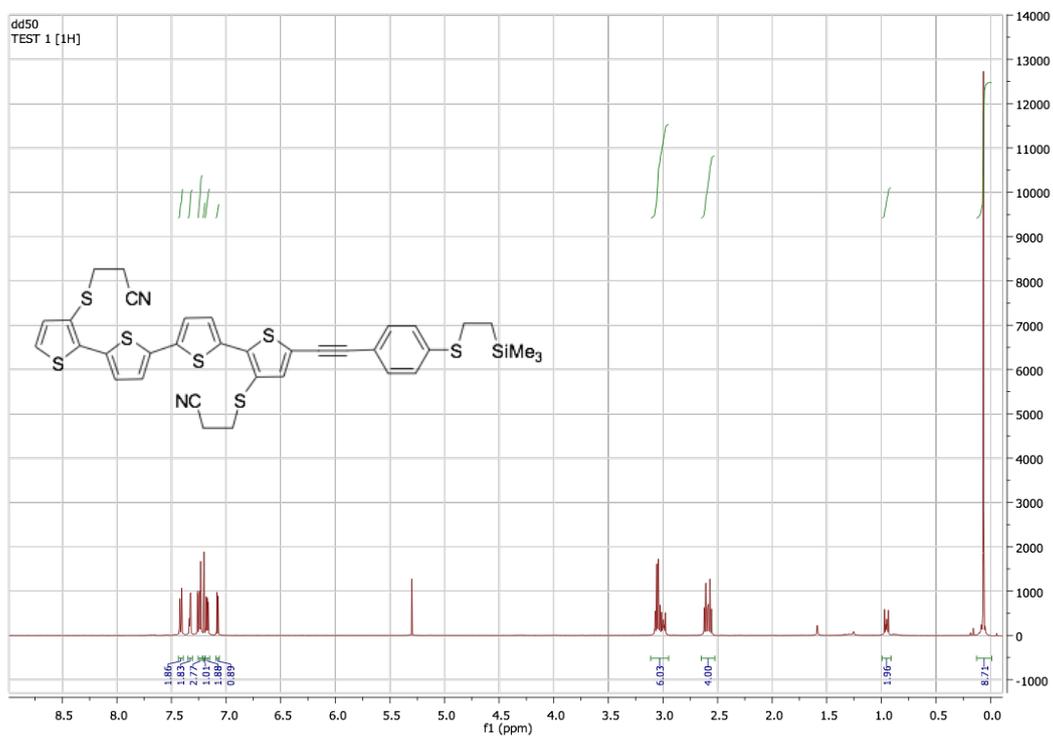

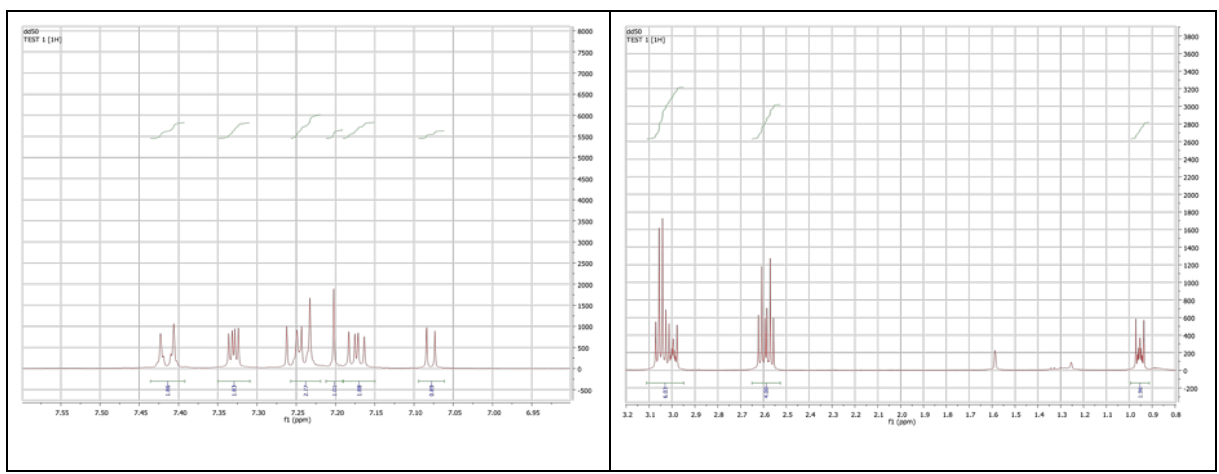

**Fig. S3.** 500 MHz $^1$H NMR spectrum of compound **4** in CDCl$_3$ at 20°C (The peak at 5.30 is assigned to traces of CH$_2$Cl$_2$).

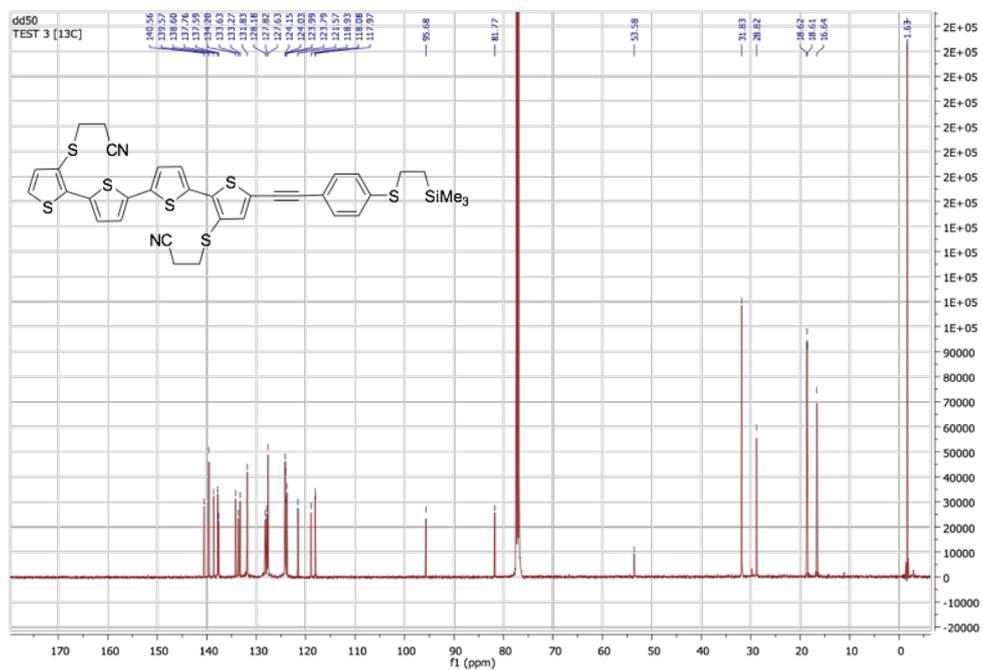

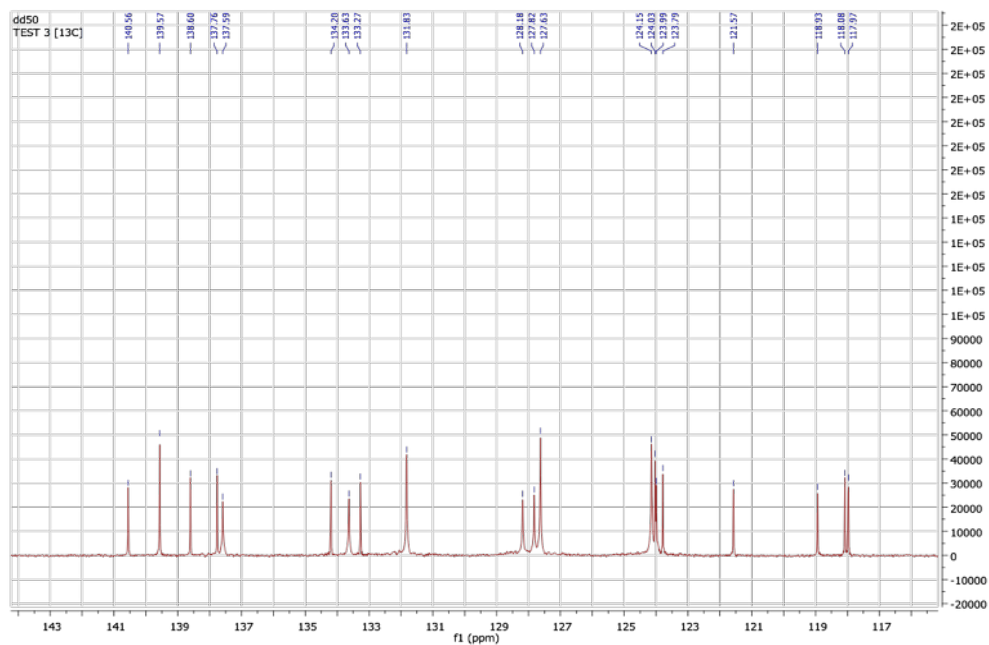

**Fig. S4.** *125 MHz $^{13}$C NMR spectrum of compound **4** in CDCl$_3$ at 20°C (The peak at 53.58 is assigned to traces of CH$_2$Cl$_2$).*

**Fig. S5.** 500 MHz [1]H NMR spectrum of compound **6** in CDCl$_3$ at 20°C.

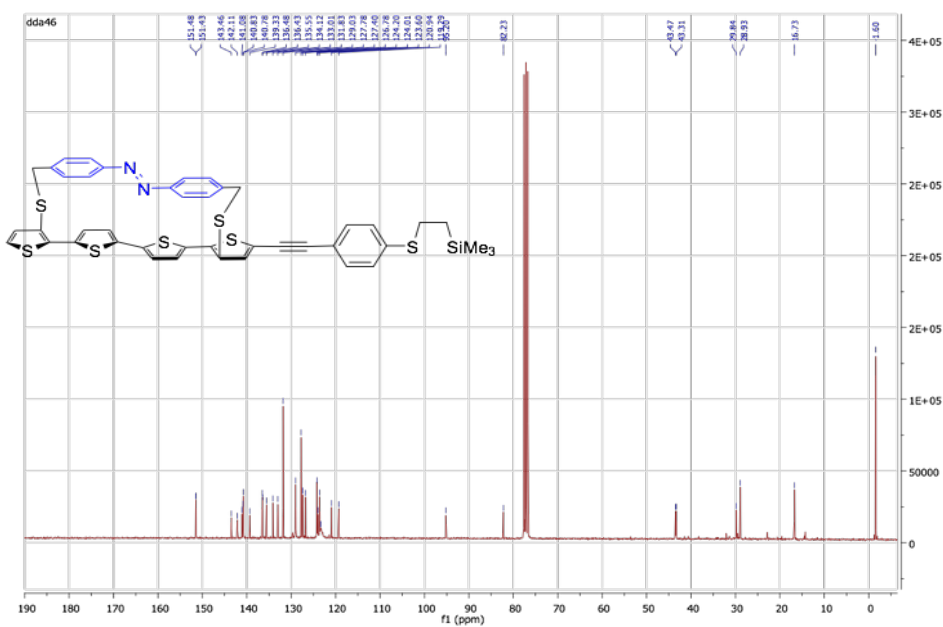

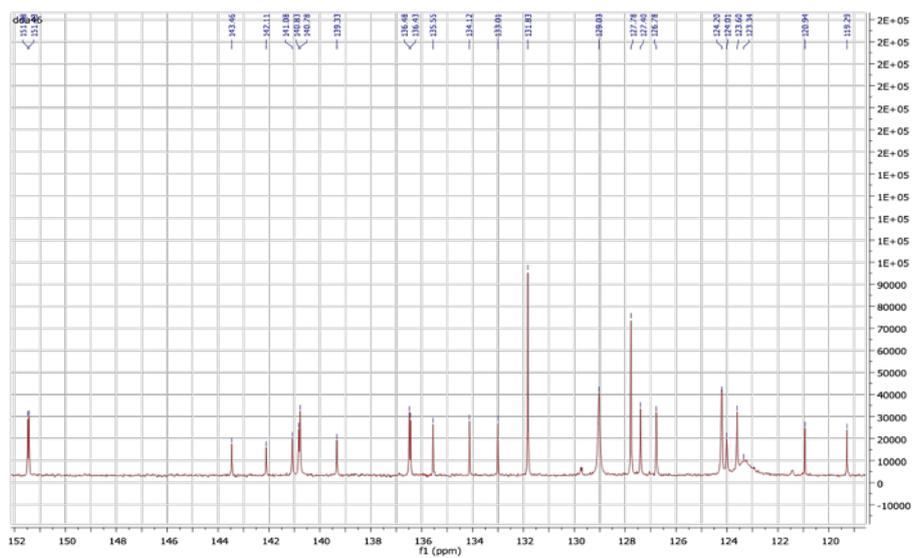

**Fig. S6.** *125 MHz $^{13}C$ NMR spectrum of compound **6** in CDCl$_3$ at 20°C (According to*

*Silverstein-Bassler-Morrill in Spectrometric identification of organic compounds, Fifth*

*edition, John Wiley & Sons, Inc., the peak at 29.84 ppm can be assigned to the presence*

*of grease).*

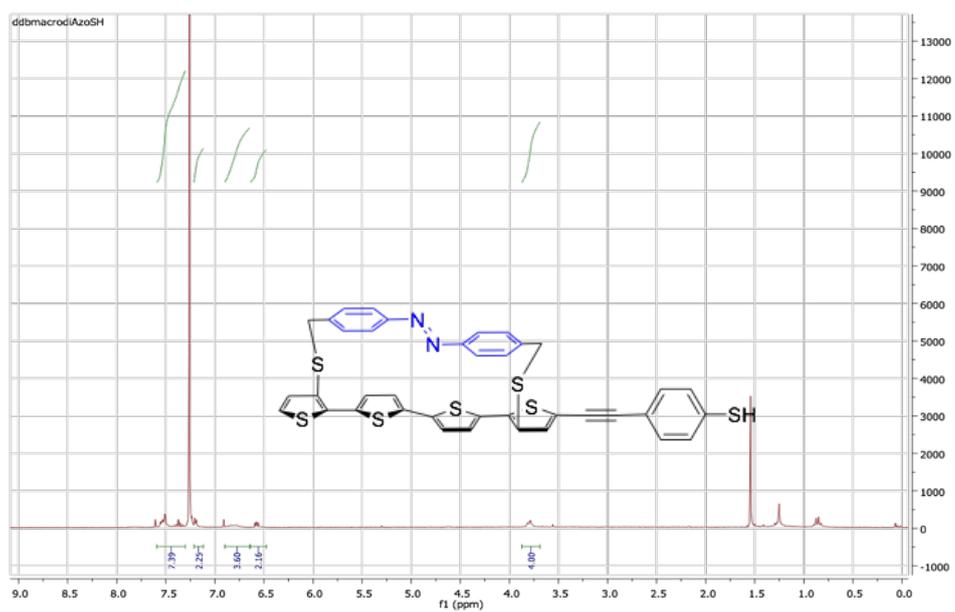

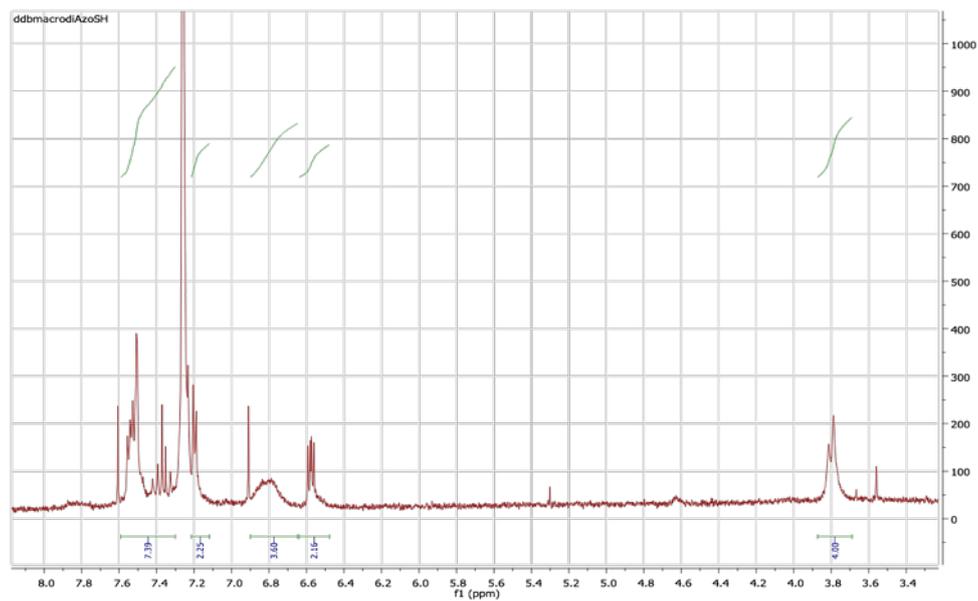

**Fig. S7.** *500 MHz* $^1H$ *NMR spectrum of compound* **7** *in CDCl$_3$ at 20°C (Due to the lack of*

*solubility in CDCl$_3$, the most intense peaks correspond to CHCl$_3$ and to H$_2$O respectively at*

*7.26 ppm and 1.55 ppm).*

## 3. Monolayer fabrication and characterization

**SAM Fabrication**: SAMS were formed on gold metallized Si wafers ⟨100⟩ n-type ($10^{-3}$ $\Omega$.cm) single side polished. Wafers were cleaned with solvents (acetone and isopropanol, VLSI grade from Carlo Erba) under ultrasonic bath and dried under nitrogen flow before deposition by sputtering of 10 nm thick titanium and 100 nm thick gold. During the deposition process wafers were heated at 350 °C in order to minimize the roughness (less 1 nm RMS). Next, the gold coated wafer was directly, after deposition, immersed in a freshly prepared millimolar solution of thiol **7** in $CH_2Cl_2$ for 1-3 days in dark and nitrogen filled glove box ($O_2$ and $H_2O$ < 1 ppm). Then, the samples were rinsed in an ultrasonic bath during 5 min with $CH_2Cl_2$ before use. We have fabricated 3 sets of samples using the above protocol but with time intervals of several months. Thus, we cannot exclude that small changes in the ambient conditions during the SAM fabrication (e.g. temperature, small changes of the hygrometry in the glove-box or quality of the solvents used, etc…), not controlled in our ordinary laboratory conditions, can result in different structure/organization of the pristine SAMs and thus different electron transport behaviors.

**Spectroscopic ellipsometry**: We recorded spectroscopic ellipsometry data in the visible range using an UVISEL (Jobin Yvon Horiba) Spectroscopic Ellipsometer equipped with a DeltaPsi 2 data analysis software. The system acquired a spectrum ranging from 2 to 4.5 eV (corresponding to 300 to 750 nm) with intervals of 0.1 eV (or 15 nm). Data were taken at an angle of incidence of 70°, and the compensator was set at 45.0°. We fitted the data by a regression analysis to a film-on-substrate model as described by their thickness and their complex refractive indexes. First, we recorded a background before monolayer deposition for the gold coated substrate. Secondly, after the monolayer deposition, we used a 2 layers model (substrate/SAM) to fit the measured data and to determine the SAM thickness. We used the previously measured optical properties of the gold coated substrate (background), and we fixed the refractive index of the organic monolayer at 1.50. The usual values in the literature for the refractive index of organic monolayers are in the range 1.45-1.50.[3-4] We can notice that a change from 1.50 to 1.55 would result in less than 1 Å error for a thickness less than 30 Å. We estimated the accuracy of the SAM thickness measurements at ± 2 Å.

**XPS measurements**: X-ray Photoelectron Spectroscopy (XPS) experiments were performed to analyze the chemical composition of the SAMs and to detect any unremoved contaminant. We used a Physical Electronics 5600 spectrometer fitted in an UHV chamber with a residual pressure of $2 \times 10^{-10}$ Torr. High resolution spectra were recorded with a monochromatic $Al_{K\alpha}$ X-ray source ($h\upsilon$=1486.6 eV), a detection angle of 45° as referenced to the sample surface, an analyzer entrance slit width of 400 µm and with an analyzer pass energy of 12 eV. In these conditions, the overall resolution as measured from the full-width half-maximum (FWHM) of the Ag $3d_{5/2}$ line is 0.55 eV. Semi-quantitative analysis were completed after standard background subtraction according to Shirley's method.[5] Peaks were decomposed by using Voigt functions and a least-square minimization procedure and by keeping constant the Gaussian and Lorentzian broadenings for each component of a given peak.

**Contact-angle measurements:** We measured the water contact angle with a remote-computer controlled goniometer system (DIGIDROP by GBX, France). We deposited a drop (10-30 μL) of desionized water (18 MΩ.cm$^{-1}$) on the surface and the projected image was acquired and stored by the computer. Contact angles were extracted by contrast contour image analysis software. These angles were determined few seconds after application of the drop. We typically measures 3-4 drops at different places on the surface and reported the average values (Fig. 6 in the main text). The error for these measurements are between ±1° and ±3° as the results of the intrinsic precision of the camera and precision of the image contour extraction.

**Light exposure**: An optical fiber was brought close to the surface. For the light irradiation, we focused the light from a xenon lamp (150 W) to the optical fiber, and we used a dichroic filter centered at 480 nm (ref. 480FS10-50 from LOT Oriel) and 360 nm (ref. 360FS10-50 from LOT Oriel) with a bandwidth of 10 nm respectively for the blue light and UV light illumination. At the output of the optical filter, the surfaces were irradiated on about 1 cm$^2$ at power density of ~70 μW/cm$^2$ at 360 nm and ~250 μW/cm$^2$ at 480 nm.

**Electrical transport with eGaIn drop contact**: To form molecular junctions, we used eutectic gallium indium drop contact (eGaIn 99.99%, Ga:In; 75.5:24.5 wt % from Alfa Aesar). We used a method close to the one developed by Chiechi et al. [6] We formed a drop of eGaIn at the extremity of a needle (with an inside diameter of 0.5 mm). The eGaIn drop volume is controlled by connecting a tubing pump (ISM832A from ISMATEC) at the extremity of the needle via standard tubing. By moving up the sample with a lab-lift, we brought gently the eGaIn drop into contact with a sacrificial surface, and we retracted the surface slowly. By this technique, we formed a conical tip of eGaIn with a diameter of around 200 μm at the extremity (corresponding to contact area of ~10$^{-3}$

cm$^2$). This conical tip was then put into contact with the SAM under control with a digital video camera. We used the eGaIn tip at different places on the samples, and we regularly formed a new tip for the same sample to avoid pollution of the tip. Note that the eGaIn tips were lifted from the surface for the light illuminations. The eutectic GaIn drop allows a good electric contact with the SAM because eGaIn accommodates well the surface roughness and can be used on gold without forming amalgam.[6] Current voltages (I-Vs) curves were acquired with a semiconductor parameters analyzer Keysight 4156C following the voltage sweep sequences 0 to +1 V and the 0 to -1V repeated several times.

## 4. Theory

**Isolated molecules:** Ab initio calculations were performed in the density functional theory framework using the Gaussian09 software. The gradient corrected functional B3LYP with a 6-311g basis set were used for all atoms. The geometry of the different isomers have been built using the experimental crystallographic structure form.[7] Different isomers (*sas*), (*aaa*), (*ssa*), (*aas*) and (*asa*) have been tested for the 4T part in the *cis* and *trans* configurations but five configurations (*cis-asa, cis-aas, cis-sas*) and (*trans-sas, trans-ssa*) exhibited a larger stability of the isomer. A local or global minimum has been observed in these different cases.

**Molecular junctions:** The five optimized structures are contacted to a first electrode, assuming a tilt of the phenyl plane of 55°. The sulfur atom lies on-top of a gold atom of a (111) surface made of 4 x 5 x 5 gold atoms. The lattice vectors of the surface unit cell are in Å (14.4179, 0.0) and (-7.20895, 9.98902). We choose a distance between the top gold surface and the upmost hydrogen atom of the molecule that is equal to the sum of the hydrogen and gold van Der Waals radii, i.e., 2.86 Å.

This scattering region is then contacted to two semi-infinite electrodes. The electronic structure and transmission spectrum of this two-probe system is computed with the NEGF-DFT procedure as implemented in the ATK2008.10 quantum-chemistry package.[8] The exchange-correlation is approximated at the GGA.revPBE level of theory.[9-10] We use a DoubleZeta+Polarization basis set for all atoms of the molecule and a SingleZeta+Polarization basis set for all gold atoms. The k-sampling for the electronic structure is 7x7x200. The k-sampling for the transmission spectrum is 5x5. The mesh cut-off for the Poisson equation solving is 300 Ry.

## 5. I-V curves for a 3[rd] set of samples

Fig. S8 shows the I-V curves recorded on a 3[rd] set of samples at different location on the SAM surfaces. Fig. S8-a shows all the I-V curves for the 4T//azo molecules in the *trans* isomer (after blue light irradiation, blue curves) and in the *cis* isomers after UV-light irradiation (purples curves). Based on the curves shown in the main text (Fig. 7) and the possible identification done by comparison with the theoretical calculations and following the same approach as described in the main text, we have marked the ascribed isomers for the different "family" of I-V curves. The situation is clearly a mix of the observed results shown in Fig. 7. The curves are also shown on two "splits" (Fig. S8-b and Fig. S8-c) for the *trans* and *cis* isomers, respectively. On these two figures we have also identified the different zones (labeled Z3 to Z8) where the corresponding I-V have been measured (two zones, Z1 and Z2, are not shown because we only measured noise in the voltage range -1V to +1V).

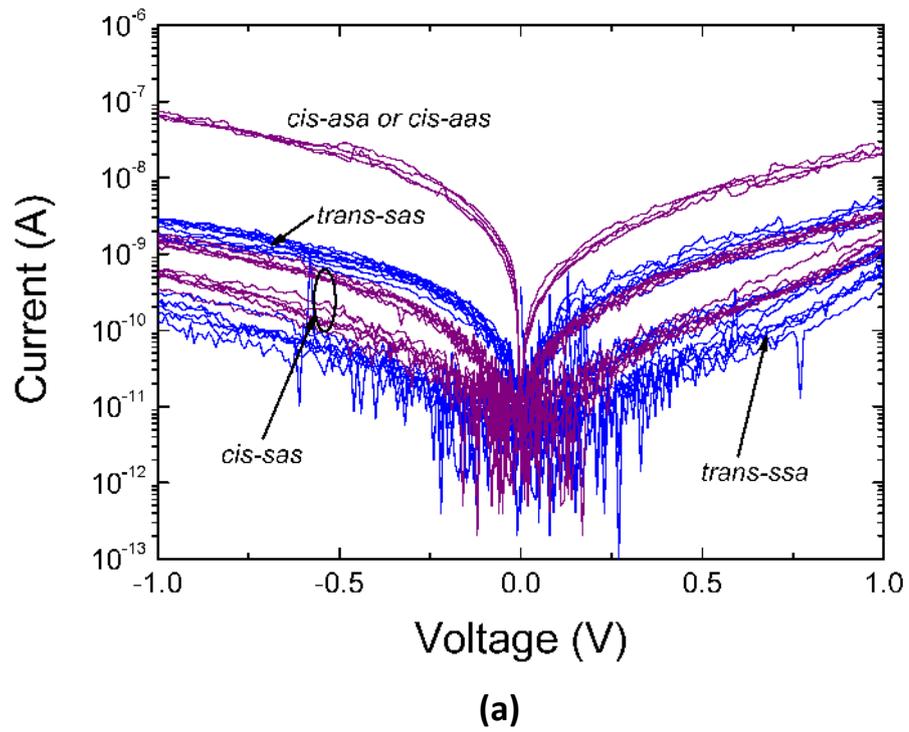

**(a)**

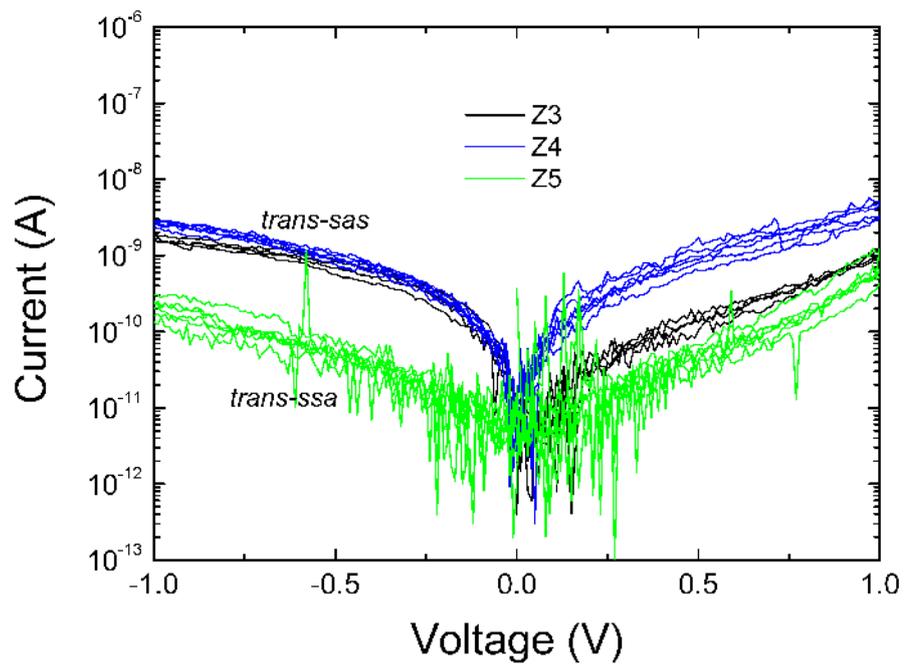

(b)

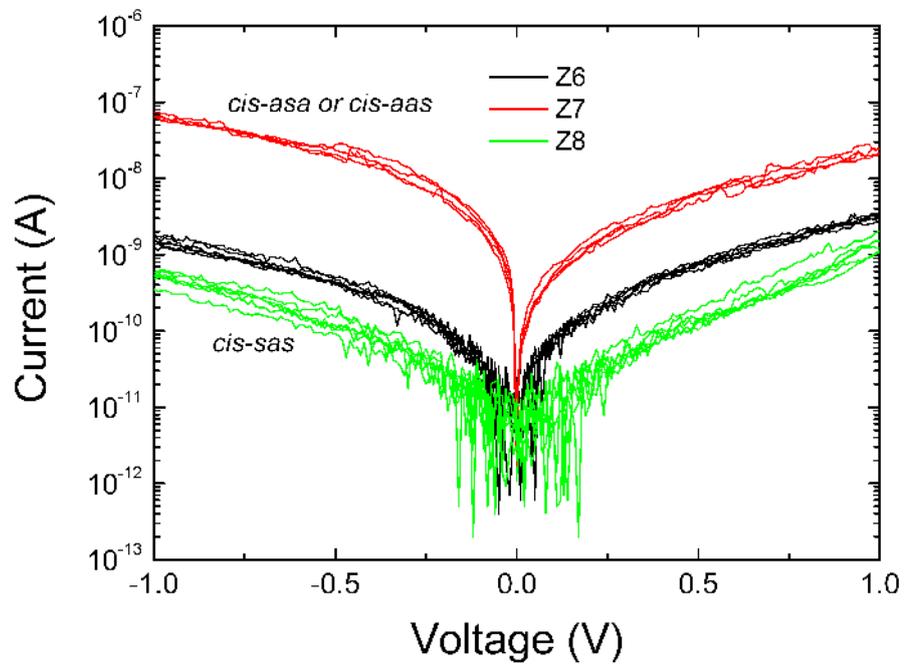

(c)

**Fig. S8. (a)** *I-V curves for a 3<sup>rd</sup> set of samples with the 4T//azo in the trans (blue curves) and cis isomer (purple curves). (b) I-V curves for the trans state measured at different zones (Z3-Z4) on the surface (identified by colors), (c) I-V curves for the cis state measured at different zones (Z6-Z8) on the surface (identified by colors). Note that there is no direct correspondence between the zones for the measurements in the trans and cis form since the eGaIn drop was lifted for light illumination and we cannot rigorously control the positioning of the drop with our system.*